\documentclass[12pt,preprint]{aastex}

\usepackage{color}

\slugcomment{}

\shorttitle{} 
\shortauthors{Moerchen et al.}

\begin{document}

\title{High-spatial-resolution imaging of thermal emission \\ from debris disks}

\author{Margaret M. Moerchen\altaffilmark{1,2}}

\altaffiltext{1}{European Southern Observatory, Alonso de C\'ordova 3107, Santiago, Chile}

\email{mmoerche@eso.org}

\author{Charles M. Telesco\altaffilmark{2} and Christopher Packham\altaffilmark{2}}

\altaffiltext{2}{University of Florida, Department of Astronomy, 211 Bryant Space Science Center, Gainesville, FL  32611, USA}

\begin{abstract}

	We have obtained sub-arcsec mid-IR images of a sample of debris disks within 100~pc.  
For our sample of nineteen A-type debris disk candidates chosen for their IR excess, we have resolved, for the first time, five sources plus the previously resolved disk around HD~141569.  Two other sources in our sample have been ruled out as debris disks since the time of sample selection. 
	Three of the six resolved sources have inferred radii of 1--4~AU (HD~38678, HD~71155, and HD~181869), and one source has an inferred radius $\sim$10--30~AU (HD~141569).  Among the resolved sources with detections of excess IR emission, HD~71155 appears to be comparable in size (r$\sim$2~AU) to the solar system's asteroid belt, thus joining $\zeta$~Lep (HD~38678, reported previously) to comprise the only two resolved sources of that class.  
	Two additional sources (HD~95418 and HD~139006) show spatial extent that implies disk radii of $\sim$1--3~AU, although the excess IR fluxes are not formally detected with better than 2-$\sigma$ significance.
	For the unresolved sources, the upper limits on the maximum radii of mid-IR disk emission are in the range $\sim$1--20~AU, four of which are comparable in radius to the asteroid belt.  We have compared the global color temperatures of the dust to that expected for the dust in radiative equilibrium at the distances corresponding to the observed sizes or limits on the sizes.  In most cases, the temperatures estimated via these two methods are comparable, and therefore, we see a generally consistent picture of the inferred morphology and the global mid-IR emission.  
		Finally, while our sample size is not statistically significant, we notice that the older sources ($>$200~Myr) host much warmer dust (T$\gtrsim$400~K) than younger sources (in the 10s of Myr).  
		%If this result were to be supported by statistics from larger samples in the future, it might indicate that asteroid-belt-type disks are more likely to be sustained at a significant level for a longer period of time than Kuiper-Belt-type disks.

\end{abstract}


\section{The Search for Resolved Disks}

Studying the structure of circumstellar debris disks is proving to be a valuable technique for inferring the presence of planets and understanding physical processes associated with the early evolution of planetary systems (see Wyatt 2008 for review)\nocite{wya08}.  The detection of an infrared excess for a main-sequence star is a strong indication that a debris disk is present.  
Through relatively large IR-photometric surveys, we can assess trends in excess IR luminosity for samples of debris disks that span a large range of ages (e.g., Su et al. 2006\nocite{su06}).  By going a step further to image single sources with high spatial resolution (with currently available facilities, $<$0.5'' is considered ``high''), we may discover structures in the dust disk that are indicative of the physical processes occurring in them.

There are presently several hundred photometric detections of debris disks (e.g., Oudmaijer et al. 1992; Mannings \& Barlow 1998; Zuckerman \& Song 2004\nocite{oud92,man98,zuc04}), but only a small number ($\sim$20) have been spatially resolved. Debris disks can be imaged in scattered light at optical or near-IR wavelengths, but such observations suffer from strong photospheric contamination from the central star, and a coronagraph typically must be employed (e.g., Weinberger et al. 1999)\nocite{wei99}. This problem is largely avoided by observing at mid-IR wavelengths where there is relatively less emission from the photosphere, and thermal dust emission dominates.  

Recent observations by \citet{cur08} demonstrate the expected decay in disk brightness with age.  That work supports the models of \citet{ken04b}, in which the disk luminosity rises sharply before slowly declining, with the peak luminosity occurring at $\sim$10--15~Myr.  Although such general trends in disk brightness with age can be inferred through survey photometric observations, the location of the dust and the physical processes that sculpt the disks are poorly understood.  For example, HR~4796A and $\beta$~Pic, both A stars, have nearly the same 18~$\mu$m/25~$\mu$m color \citep{moe07b}, and \citet{wya08} notes that they occupy similar positions in the age-vs.-24-$\mu$m-excess plot presented by \citet{cur08}.  Despite such apparent similarities, resolved near-IR and mid-IR images of both sources reveal significantly different dust distributions: HR~4796A has a well-defined dust annulus at $\sim$70~AU that is 17~AU wide \citep{sch99, wya99, tel00} and the dominant mid-IR-emitting dust disk of Beta Pic spans 20-120 AU \citep{lag94,tel05} (although optically scattered light reveals a disk extending out to nearly 1500 AU [e.g., Larwood \& Kalas 2001]).  Thus, the excess flux levels tell only a small part of the story, but  degeneracies such as the example above can sometimes be broken through imaging observations.

%% since they only sample dust particles that are heated to $\sim$150--1500~K, 

The initial goal of this research was to explore how morphological asymmetries (or lack thereof) are generated by various physical processes such as collisions and orbital resonances.  Imaging at several wavelengths permits some assessment of which process may dominate a particular disk or region therein.  The mid-IR regime offers the additional benefit of near-diffraction-limited observing, and by imaging from the ground, as in this work, we can exploit large telescopes to achieve the desired high angular resolution.   For example, the $\lambda/D$ diffraction limits at 11.7~$\mu$m and 18.3~$\mu$m at the 7.9-meter Gemini Observatory telescopes are 0.24" and 0.39", respectively.

\section{Observations}

\subsection{Source Sample
 \label{sec:sample}
}

	Seventeen of the nineteen debris disk candidates observed in this program are associated with the stars listed in Table~\ref{tab:targets}.  They are main-sequence stars (with the exception of HD~141569), essentially A-type (B8--A5), all within 100~pc.  In the literature, HD~141569 is considered to be a transition disk, since a significant amount of gas (e.g., CO) has been detected within it \citep{bri02, bri03}.  All sources were observed either with MIPS on $Spitzer$ at 24~$\mu$m or with $IRAS$ at 25~$\mu$m.  The sources in this sample were chosen both for their high disk-to-star ratio of excess emission ($>$1.1) at 24~$\mu$m \citep{rie05} as well as their high estimated flux densities at 10~$\mu$m ($>$10~mJy) and 18~$\mu$m ($>$40~mJy) attributable to dust emission.  Two sources, HD~172555 and HD~181296, were chosen for their $IRAS$-discovered high fractional dust luminosities \citep{oud92, man98} that were confirmed with MIPS and/or ISO \citep{moo06}.  The ages of the sample stars are in the range $\sim$5--600~Myr.  Since the time of sample selection, other works have demonstrated that two of the targets that we observed (not listed in Table~1) have infrared excesses that cannot be attributed to debris disk processes.  These instances are reviewed individually in this section, and will not be discussed in future sections regarding debris disk analysis. 
	
\begin{deluxetable}{lcllcl}
\tablecolumns{6}
\tablewidth{0pc} 
\tablecaption{Summary of Imaging Observations\label{tab:observations}}
\tablehead{
\colhead{Object} & \colhead{Gemini} & \colhead{Program ID} & \colhead{Filter} & \colhead{Time$^{a}$ [s]} & \colhead{Dates Observed}
} 
\startdata
%HD 21362 & N & 2006A-Q-10 & N$^\prime$ & 10 Sep 2006, 22 Sep 2006 \\
% &  &  & Qa & 13 Oct 2006 \\
% &  &  & Qa & 24 Jan 2006, 3 Feb 2006, 4 Mar 2006, 9~Mar~2006, 10~Mar~2006 \\
HD 38206 & S & 2005A-Q-2 & N & 900 & 19 Sep 2005 \\
 &  &  & Qa & 900 & 5 Feb 2006 \\
HD 38678 & S & 2005A-Q-2 & N & 900 & 3 Feb 2005 \\
 &  &  & Qa & 900 & 3 Feb 2005 \\
HD 56537 & N & 2006A-Q-10 & N$^\prime$ & 900 & 4 Apr 2006 \\
 &  &  & Qa & 900 & 4 Apr 2006, 7 Apr 2006 \\
HD 71155 & S & 2005A-Q-2 & N & 900 & 4 Mar 2006 \\
 &  &  & Qa & 900 & 5 Feb 2006 \\
%HD 74956 & S & 2005A-Q-2 & N & & 6 Feb 2006 \\
% &  &  & Qa & & 5 Feb 2006 \\
HD 75416 & S & 2005A-Q-2 & Si-5 & 300 & 22 May 2005 \\
 &  &  & Qa & 900 & 22 May 2005 \\
HD 80950 & S & 2005A-Q-2 & N & 900 & 8 Mar 2006 \\
 &  &  & Qa & 900 & 5 Feb 2006 \\
HD 83808 & N & 2006A-Q-10 & N$^\prime$ & 900 & 4 Apr 2006 \\
 &  &  & Qa & 900 & 6 Apr 2006 \\
HD 95418 & N & 2006A-Q-10 & N$^\prime$ & 900 & 10 Jun 2006 \\
 &  &  & Qa & 900 & 7 Apr 2006 \\
HD 102647 & N & 2006A-Q-10 & N$^\prime$ & 960 & 11 Jun 2006 \\
 &  &  & Qa & 900 & 10 May 2006, 15 May 2006 \\
HD 115892 & S & 2005A-Q-2 & 
%N & 900 & 21 May 2005 \\
% &  &  & 
 Si-5 & 900 & 22 May 2005 \\
 &  &  & Qa & 1500 & 21 May 2005, 22 May 2005 \\
HD 139006 & N & 2006A-Q-10 & N$^\prime$ & 900 & 29 May 2006 \\
 &  &  & Qa & 900 & 30 Apr 2006, 12 May 2006 \\
HD 141569 & N & 2006A-Q-10 & N$^\prime$ & 900 & 12 Jun 2006 \\
 &  &  & Qa & 900 & 10 May 2006, 14 May 2006 \\
HD 161868 & N & 2006A-Q-10 & N$^\prime$ & 900 & 29 May 2006 \\
 &  &  & Qa & 900 & 30 Apr 2006, 20 May 2006 \\
HD 172555 & S & 2007A-Q-23 & Si-5 & 680 & 1 Jul 2007 \\
 &  &  & Qa & 680 & 27 Jun 2007, 1 Jul 2007 \\
HD 178253 & S & 2005A-Q-2 & Si-5 & 900 & 22 May 2005 \\
 &  &  & Qa & 600 & 22 May 2005 \\
HD 181296 & S & 2007A-Q-23 & Si-5 & 700 & 28 Apr 2007 \\
 &  &  & Qa & 700 & 28 Apr 2007 \\
HD 181869 & S & 2005A-Q-2 & N & 900 & 20 Aug 2005 \\
 &  &  & Qa & 1200 & 20 Aug 2005, 19 Mar 2006 \\
\enddata
\\	\small{Notes-- $^a$Time is on-source integration time.}

\end{deluxetable}

\textbf{HD~21362}-- The IRS (Infrared Spectrograph) instrument on $Spitzer$ obtained low-resolution spectra of HD~21362 that showed several hydrogen emission lines that are indicative of free-free radiation from an ionized stellar wind.  \citet{su06} concluded that the infrared excess previously thought to be due to a debris disk presence is actually due to a fast-rotating B-type star with a strong stellar wind creating a circumstellar gas disk (also known as the Be phenomenon.)  HD~21362 is not resolved in our images, and its measured flux densities are 683~$\pm$~68~mJy at 11.2~$\mu$m and 456~$\pm$~68~mJy at 18.1~$\mu$m.

\textbf{HD 74956}-- Recent $Spitzer$ MIPS images at 24~$\mu$m have demonstrated that the observed infrared excess associated with HD~74956 (e.g., Aumann 1985\nocite{aum85}, 1988\nocite{aum88}; Cote 1987\nocite{cot87}; Chen et al. 2006\nocite{che06}; Su et al. 2006\nocite{su06}) is the result of this multiple-star system traveling through an interstellar cloud and producing a bow shock \citep{gas08}.  The dust in this overdense region ($\sim$15 times the Local Bubble density) is compressed at the shock front generated by photon pressure and is heated by the star, which gives rise to an arc-shaped morphology that is responsible for the infrared excess.   More recently, \citet{ker09} studied the system in greater detail at both near-IR and mid-IR wavelengths (with NACO and VISIR, respectively, at the VLT) to determine whether some of the observed excess might still be attributable to a disk around one of the system members.  However, their final result corroborates that of \citet{gas08}: the bow shock alone is likely to be responsible for the IR excess.  HD~74956 is resolved only in our 10.4-$\mu$m images, and the level of extension implies a dust disk radius of 1.4~AU (simply estimated by quadratically subtracting the PSF FWHM from the source FWHM).  The measured flux densities are 8.61~$\pm$~0.86~mJy at 10.4~$\mu$m and 2.38~$\pm$~0.36~mJy at 10.4~$\mu$m

\begin{deluxetable}{lccccccc}
\tablecolumns{8}
\tablewidth{0pc} 
\tablecaption{Debris Disk Candidate Target List
\label{tab:targets}}
\tablehead{
\colhead{Name} & \colhead{Type} & \colhead{Age} & \colhead{Age} & \colhead{$d$} & \colhead{24 or 25um $F_{\nu}$} & \colhead{Flux} & \colhead{Excess Ratio\tablenotemark{a}} \\
\colhead{} & \colhead{} & \colhead{[Myr]} & \colhead{Ref.}  & \colhead{[pc]} & \colhead{[Jy]} & \colhead{Ref.} & \colhead{[$F_{total}/F_{\star}$]}} 
 \startdata
%HD 21362 & B6Vn & 80 & 1 & 170 & 0.324 & 4 & 8.37 \\
HD 38206 & A0V & 9 & 2 & 69 & 0.115 & 4 & 3.34 \\
HD 38678 & A2Vann & 231, 330 & 1, 3 & 22 & 1.160 & 9 & 2.43 \\
HD 56537 & A3V & 560 & 3 & 29 & 0.586 & 9 & 1.32 \\
HD 71155 & A0V & 169, 240 & 1, 3 & 38 & 0.321 & 4 & 1.54 \\
%HD 74956 & A1V & 390, 330 & 1, 4 & 24 & 1.990 & 12 & 1.1 \\
HD 75416 & B8V & 5 & 4 & 97 & 0.128 & 4 & 3.51 \\
HD 80950 & A0V & 80 & 2 & 81 & 0.121 & 4 & 3.79 \\
HD 83808 & A5V+ & 400 & 3 & 41 & 1.140 & 9 & 1.16 \\
HD 95418 & A1V & 300, 358, 380 & 5, 1, 3 & 24 & 1.400 & 9 & 1.21 \\
HD 102647 & A3V & 50, 520 & 1, 3 & 11 & 2.320 & 9 & 1.42 \\
HD 115892 & A2V & 350 & 3 & 18 & 0.705 & 4 & 1.2 \\
HD 139006 & A0V & 314, 350 & 1, 3 & 23 & 1.686 & 9 & 1.29 \\
HD 141569 & B9.5e & 5 (PMS) & 6, 7 & 99 & 1.819 & 9 & 162.4\tablenotemark{b} \\
HD 161868 & A0V & 184, 305 & 1, 3 & 29 & 0.525 & 9 & 1.47 \\
HD 172555 & A5IV-V & 12 & 8 & 29 & 1.092 & 9 & 9.66\tablenotemark{b} \\
HD 178253 & A2V & 254, 320 & 1, 3 & 40 & 0.348 & 9 & 1.45\tablenotemark{b} \\
HD 181296 & A0Vn & 12 & 8 & 48 & 0.491 & 9 & 8.32 \\
HD 181869 & B8V & 110 & 3 & 52 & 0.280 & 9 & 1.46 \\
\enddata
	\tablenotetext{a}{For flux density measurements at 24 or 25~$\mu$m.}
	\tablenotetext{b}{These excess ratios were computed with $IRAS$ 25~$\mu$m flux density measurements and 25~$\mu$m photospheric flux densities estimated with the same method as described in \S\ref{sec:phot_estimates}.  All other excess ratio values are taken from Rieke et al. 2005.}
	\tablerefs{
	(1) Song et al. 2001\nocite{son01}; 
%	(2) Kalas 2005\nocite{kal05};
	(2) Gerbaldi et al. 1999\nocite{ger99};
	(3) Rieke et al. 2005\nocite{rie05};
	(4) de Zeeuw et al. 1999\nocite{dez99};
	(5) King et al. 2003\nocite{kin03};
	(6) Weinberger et al. 2000\nocite{wei00};
	(7) Mer\'in et al. 2004\nocite{mer04};
	(8) Zuckerman et al. 2001\nocite{zuc01};
%	(10) Stauffer et al. 1995\nocite{sta95};
	(9) Moshir et al. 1989\nocite{mos89}
%	(10) IPAC 1986\nocite{ipa86}
	}
	
\end{deluxetable}

\subsection{The Images: General Comments} \label{sec:data}

	We obtained mid-IR images of 19 debris disk candidates (including HD~21362 and HD~74956 mentioned above) in 2005, 2006, and 2007 at the Gemini North and South facilities (program IDs: GS-2005A-Q-2, GN-2006A-Q-10, GS-2006A-Q-5, GS-2007A-Q-23) with Michelle and T-ReCS (Thermal Region Camera and Spectrograph), respectively.  The log of the observation dates and program IDs is given in Table~\ref{tab:observations} in the appendix. 
	With Michelle, we used the narrowband N$^\prime$ ($\lambda_c$~=~11.2~$\mu$m, $\Delta\lambda$~=~2.4~$\mu$m) and narrowband Qa ($\lambda_c$~=~18.1~$\mu$m, $\Delta\lambda$~=~1.9~$\mu$m) filters.  
	With T-ReCS, we used the broadband N ($\lambda_c$~=~10.36~$\mu$m, $\Delta\lambda$~=~5.27~$\mu$m), narrowband Si-5 ($\lambda_c$~=~11.66~$\mu$m, $\Delta\lambda$~=~1.13~$\mu$m), and narrowband Qa ($\lambda_c$~=~18.30~$\mu$m, $\Delta\lambda$~=~1.51~$\mu$m) filters.   These filters were chosen to sample the disk emission in the two atmospheric transmission windows in the mid-IR regime, at $\sim$10 (N band) and $\sim$20~$\mu$m (Q band).  Among the filters available in the N band, the N$^\prime$ filter in Michelle and the broadband N and Si-5 filters in T-ReCS have the best sensitivity, as documented in Gemini-provided tables.  The Qa filter ($\sim$18~$\mu$m) was chosen for its relative lack of water absorption lines in its wavelength range compared to Qb filters at $\sim$25~$\mu$m, thus decreasing the importance of low atmospheric water vapor content for execution of the observations.   
	 Both Michelle and T-ReCS utilize Raytheon Si:As blocked-impurity-band (BIB) detectors with 320 x 240 pixels.  With Michelle, each pixel subtends 0.10", and the total field of view is 32'' x 24", and with T-ReCS, each pixel subtends 0.09", and the total field of view is 29" x 22".   
	 
	 A point-spread-function (PSF) comparison star was observed before and after each science target observation, with three exceptions where the PSF star was only observed either before or after (but not both) the disk target.  The PSF reference star was typically a Cohen IR standard \citep{coh99} that also served as a flux calibrator.  The observations used the standard mid-IR technique of chopping with a chop throw of 15'' and nodding (parallel to the chopping direction) to remove time-variable sky background, telescope thermal emission, and low-frequency detector noise. The data were reduced with the Gemini IRAF package. 
	 	 
	   The total (disk + star) flux densities of the sources observed in this study are given in Table \ref{tab:fluxmeas}.  The flux densities were measured with aperture photometry.  The average sky background was measured in an annulus centered on the star with a radius range of 1.5--2 times the main (source) aperture radius, and the measured source flux density was corrected for the sky according to the number of pixels in the aperture.   Observations of the flux standards were not repeated on each night, so we adopt nominal calibration uncertainties of 10\% at 11.7~$\mu$m and 15\% at 18.3~$\mu$m, which are typical for photometric variations in the mid-IR (e.g. De Buizer et al. 2005, Packham et al. 2005).   Such variations dominate the uncertainties associated with the background shot noise in all cases.  The 1-$\sigma$ uncertainties presented in Table~\ref{tab:fluxmeas} represent the dispersion in the measurements due to fluctuations in the level of thermal emission of the sky and the shot noise in the background photon stream.  The total 1-$\sigma$ uncertainties (the quadratic addition of both background and photometric uncertainties) are given in Tables \ref{tab:excessm} and \ref{tab:excesst}.

\begin{deluxetable}{lccccccccccc}
\tablecaption{
Flux Densities\tablenotemark{a} (in mJy) of Debris Disk Candidates\label{tab:fluxmeas}
}
\tablecolumns{7}
\tablewidth{0pc} 
\tablehead{
\colhead{Source} & \colhead{Gemini} & \colhead{$F_{\nu}(10.4~\mu$m)}  & \colhead{$F_{\nu}(11.2~\mu$m)} & \colhead{$F_{\nu}(11.7~\mu$m)} & \colhead{$F_{\nu}(18.1~\mu$m)} & \colhead{$F_{\nu}(18.3~\mu$m)}  \\
\colhead{HD \#} & \colhead{} & \colhead{(N)} & \colhead{(N$^{\prime}$)} & \colhead{(Si-5)} & \colhead{(Qa)} & \colhead{(Qa)} 
}
\startdata
%21362 & N &  ... & 683 ~$\pm$~2 &   &  456 ~$\pm$~6 & ... \\
38206 & S & 202  ~$\pm$~1 & ...  & ... & ... &    116 ~$\pm$~7 \\
38678 & S & 2147  ~$\pm$~2 &  ...  & ... & ...  &   960 ~$\pm$~10 \\
56537 & N &   & 1432  ~$\pm$~1 & ...   & 597 ~$\pm$~12 & ... \\
71155 & S & 1083  ~$\pm$~2 &  ...  & ... & ...   &  375 ~$\pm$~10 \\
%74956 & S & 8610  ~$\pm$~4 & ...   & ... &  ...  &  2380 ~$\pm$~24 \\
75416 & S & ...  & ... & 229 ~$\pm$~3 & ... &    92 ~$\pm$~7 \\
80950 & S & 191  ~$\pm$~13 &  & ...  & ...  &  109 ~$\pm$~10 \\
83808 & N & ...  & 3614 ~$\pm$~3 &  ...  & 1567 ~$\pm$~13 & ... \\
95418 & N &  ...  & 3588 ~$\pm$~2 & ...  & 1743 ~$\pm$~10 & ... \\
102647 & N &  ...  & 5822 ~$\pm$~3 &  ...  & 2317 ~$\pm$~17  & ... \\
115892 & S & 2540 ~$\pm$~3 & ...   & 2521 ~$\pm$~2 & ... &  1026 ~$\pm$~17 \\
139006 & N &  ... & 4077 ~$\pm$~2 &  ...  & 1795 ~$\pm$~15  & ... \\
141569 & N &  ...  & 338 ~$\pm$~1 &  ...  & 883 ~$\pm$~13  & ... \\
161868 & N &  ...  & 1105 ~$\pm$~1 &  ...  & 443  ~$\pm$~11  & ... \\
172555 & S &  ...  & ...  & 1155 ~$\pm$~2 & ... &  1094  ~$\pm$~11 \\
178253 & S &  ...  & ...  & 770 ~$\pm$~2 & ...  &  360  ~$\pm$~12 \\
181296 & S &  ...  & ...  & 395 ~$\pm$~2 & ...  &  343 ~$\pm$~16 \\
181869 & S & 695  ~$\pm$~2 &  ...  & ... & ... &   202 ~$\pm$~14 \\
\enddata	
\tablenotetext{a}{The uncertainties given here are those associated with the measured background noise.  Additionally, nominal calibration uncertainties of 10\% at 11.7~$\mu$m and 15\% at 18.3~$\mu$m were adopted, which are typical for photometric variations in the mid-IR (e.g. De Buizer et al. 2005, Packham et al. 2005).  These photometric uncertainties dominate the background noise in all cases.}

\end{deluxetable}

\begin{deluxetable}{lccccccc}

\tablecaption{Excess Emission of Debris Disk Candidates (Michelle)\label{tab:excessm}}

\tablewidth{0pc}

\tablehead{
\colhead{} & \multicolumn{3}{c}{$F_{\nu}$(11.2~$\mu$m) [mJy]} & & \multicolumn{3}{c}{$F_{\nu}$(18.1~$\mu$m) [mJy]} \\
 \cline{2-4} \cline{6-8} \\
\colhead{HD} & \colhead{Total} & \colhead{Star} & \colhead{Excess} & & \colhead{Total} & \colhead{Star} & \colhead{Excess}  
}

\startdata
56537 &	1432 $\pm$ 143 &	1154 &	278 $\pm$ 143 & &	597	$\pm$ 91  & 440	 & 157 $\pm$ 91 \\
83808 &	3614 $\pm$ 361 &	2985 &	628 $\pm$ 361 & &	1567 $\pm$ 235	 & 1138	 & 429 $\pm$ 235 \\
95418 &	3588 $\pm$ 359 &	3650 &	-62 $\pm$ 359 & &	1743 $\pm$ 262 &	 1392 &	351 $\pm$ 262 \\
102647 &	5822 $\pm$ 582 &	5297 &	525 $\pm$ 582 &  &	2317 $\pm$ 348 &	2020 &	297 $\pm$ 348  \\ 
139006	 & 4077$\pm$ 408	 & 3924 &	153 $\pm$ 408 & &	1795 $\pm$ 262 &	1496 &	299 $\pm$ 262 \\
141569	 & 338 $\pm$ 34  &	56 &	282 $\pm$ 34 & &	 883 $\pm$ 147 &	22 & 	861 $\pm$ 147 \\
161868	 & 1105	$\pm$ 111 & 1064 &	41 $\pm$ 111 & &	443 $\pm$ 72 & 	406 &	37 $\pm$ 72 \\
\enddata
\begin{flushleft}
\small{Notes-- The 1-$\sigma$ uncertainties given here are the quadratic addition of both background and photometric uncertainties.  For the measurement uncertainties alone, see Table~\ref{tab:fluxmeas}. 
	}
\end{flushleft}

\end{deluxetable}

\begin{deluxetable}{lcccccccccccc}
\rotate

\tablecaption{Excess Emission of Debris Disk Candidates (T-ReCS)\label{tab:excesst}}

\tablewidth{0pc}

\tablehead{
\colhead{} & \multicolumn{3}{c}{$F_{\nu}$(10.4~$\mu$m) [mJy]} & & \multicolumn{3}{c}{$F_{\nu}$(11.7~$\mu$m) [mJy]} & & \multicolumn{3}{c}{$F_{\nu}$(18.3~$\mu$m) [mJy]} \\
 \cline{2-4} \cline{6-8} \cline{10-12} \\
\colhead{HD} & \colhead{Total} & \colhead{Star} & \colhead{Excess} & & \colhead{Total} & \colhead{Star} & \colhead{Excess}  & & \colhead{Total} & \colhead{Star} & \colhead{Excess}  
}

\startdata
38206 &	202 $\pm$ 20 &	169  & 	33 $\pm$ 20 &	& ...	& ...  & ... &	&	116 $\pm$ 19	 & 54 & 	62 $\pm$ 19 \\
38678 &	2130 $\pm$ 190 &	1388  & 	742 $\pm$ 190 &	& ...	& ...  & ... &	&	960 $\pm$ 60	 & 484 & 	476 $\pm$ 60 \\
71155 &	1083 $\pm$ 108 & 	815 &	268	 $\pm$ 108 & & ... & ... & ... &	&	375 $\pm$ 57	 & 261	 & 114 $\pm$ 57 \\
75416 & ... & ...	& ... 	&	&	229	$\pm$ 23 & 141 & 	88 $\pm$ 23 & & 	92	$\pm$ 15 & 58 &	34 $\pm$ 15 \\
80950 &	191 $\pm$ 23 &	 162 &	29 $\pm$ 23	 & & ... & ... & ... & &			109 $\pm$ 19 &	52 &	57 $\pm$ 19 \\
115892 &	2540 $\pm$ 254 &	2748 &	-208 $\pm$ 254 & &	2521 $\pm$ 252 &	2155	 & 366 $\pm$ 252  & &	1026 $\pm$ 155 &	881 &	145 $\pm$ 155 \\
172555 & ...	& ...	& ...	& & 	1155 $\pm$ 116 &	520 &	635 $\pm$ 116 & &	1094 $\pm$ 164  &	213 &	881 $\pm$ 164 \\
178253 & ...	& ... 	& ... 	& & 	770	$\pm$ 77 & 655  & 115 $\pm$ 77 & &	360	$\pm$ 55 & 268	 & 92 $\pm$ 55 \\
181296 & ...	& ... 	& ... 	& &	 395	 $\pm$ 40 & 271 &	124 $\pm$ 40 & & 	343	$\pm$ 54 & 111 &	232 $\pm$ 54 \\
181869 &	695 $\pm$ 70 & 	731 & 	-36 $\pm$ 70 &	& ...  & ...  & ...  &	 &		202 $\pm$ 34	 & 234	 & -33 $\pm$ 34 \\
\enddata
\begin{flushleft}
\small{Notes-- The 1-$\sigma$ uncertainties given here are the quadratic addition of both background and photometric uncertainties.  For the measurement uncertainties alone, see Table~\ref{tab:fluxmeas}. 
	}
\end{flushleft}
\end{deluxetable}

\begin{table}[]

	\caption{Summary of Debris Disk Candidate IR Excess Detections}\label{tab:excessdetections}
	\begin{tabular}{l c c c c c}
	\hline
HD & N (10.4~$\mu$m) & N$^\prime$ (11.2~$\mu$m) & Si-5 (11.7~$\mu$m) & Qa$^a$ (18.1~$\mu$m) & Qa$^b$ (18.3~$\mu$m) \\
 \hline

38206 & none & ... & ... & ... & significant \\
38678 & significant & ... & ... & ... & significant \\
56537 & ... & none & ... & none & ... \\
71155 & marginal & ... & ... & ... & marginal \\
75416 & ... & ... & significant & ... & marginal \\
80950 & none & ... & ... & ... & significant \\
83808 & ... & none & ... & none & ... \\
95418 & ... & none$^*$ & ... & none$^*$ & ... \\
102647 & ... & none & ... & none & ... \\
115892 & ... & ... & none & ... & none \\
139006 & ... & none$^*$ & ... & none$^*$ & ... \\
141569 & ... & significant & ... & significant & ... \\
161868 & ... & none & ... & none & ... \\
172555 & ... & ... & significant & ... & significant \\
178253 & ... & ... & none & ... & none \\
181296 & ... & ... & significant & ... & significant \\
181869 & none & ... & ... & ... & none \\

 \hline
	\end{tabular}
	\small{Notes-- $^a$~Michelle, Gemini North.  $^b$~T-ReCS, Gemini South.
	
	$^*$ While there is no statistically significant detection of excess emission for these sources in our data, they do appear to be spatially resolved.  This point is discussed in \S\ref{sec:phot_estimates} and \S\ref{sec:resolved}. }
	
\end{table}

\begin{deluxetable}{lccccc}

\tablecaption{FWHM of Debris Disk Candidates \& PSF Reference Stars (Michelle)\label{tab:FWHMM}}

\tablewidth{0pc}

\tablehead{
\colhead{} & \multicolumn{2}{c}{N$^{\prime}$~FWHM [arcsec]} & & \multicolumn{2}{c}{Qa~FWHM [arcsec]} \\
 \cline{2-3} \cline{5-6} \\
\colhead{Name} & \colhead{Source} & \colhead{PSF} & & \colhead{Source} & \colhead{PSF}  
}

\startdata
%HD~21362 & 0.425 ~$\pm$~ 0.046 & 0.386 ~$\pm$~ 0.004 & & 0.583 ~$\pm$~ 0.087 & 0.552 ~$\pm$~ 0.087 \\
%  &    0.364 ~$\pm$~ 0.009 & 0.386 ~$\pm$~ 0.041   &      &  &  \\
HD~56537 & 0.365 ~$\pm$~ 0.002 & 0.398 ~$\pm$~ 0.013 & & 0.545 ~$\pm$~ 0.019 & 0.539 ~$\pm$~ 0.004 \\
  &  &  & & 0.609 ~$\pm$~ 0.025 & 0.554 ~$\pm$~ 0.006 \\
HD~83808 & 0.347 ~$\pm$~ 0.001 & 0.380 ~$\pm$~ 0.009 & & 0.537 ~$\pm$~ 0.004 & 0.529 ~$\pm$~ 0.005 \\
HD~95418 & 0.339 ~$\pm$~ 0.001 & 0.328 ~$\pm$~ 0.002 & & 0.539 ~$\pm$~ 0.003 & 0.543 ~$\pm$~ 0.003 \\
HD~102647 & 0.361 ~$\pm$~ 0.002 & 0.353 ~$\pm$~ 0.008 & & 0.533 ~$\pm$~ 0.005 & 0.535 ~$\pm$~ 0.002 \\
  &      &      &  & 0.531 ~$\pm$~ 0.003 & 0.533 ~$\pm$~ 0.003 \\
HD~139006 & 0.419 ~$\pm$~ 0.003 & 0.364 ~$\pm$~ 0.006 & & 0.574 ~$\pm$~ 0.007 & 0.556 ~$\pm$~ 0.004 \\
  &      &      & &  0.516 ~$\pm$~ 0.004 & 0.544 ~$\pm$~ 0.003 \\
HD~141569 & 0.436 ~$\pm$~ 0.004 & 0.374 ~$\pm$~ 0.016 & & 0.807 ~$\pm$~ 0.029 & 0.539 ~$\pm$~ 0.005 \\
  &      &      & & 0.818 ~$\pm$~ 0.066 & 0.523 ~$\pm$~ 0.003 \\
HD~161868 & 0.386 ~$\pm$~ 0.003 & 0.356 ~$\pm$~ 0.006 & & 0.518 ~$\pm$~ 0.031 & 0.528 ~$\pm$~ 0.003 \\
  &      &      & & 0.537 ~$\pm$~ 0.017 & 0.525 ~$\pm$~ 0.006 \\
\enddata

\end{deluxetable}

\begin{deluxetable}{lcccccccc}

\rotate

\tablecaption{FWHM of Debris Disk Candidates \& PSF Reference Stars (T-ReCS)\label{tab:FWHMT}}

\tablewidth{0pc}

\tablehead{
\colhead{} & \multicolumn{2}{c}{N~FWHM [arcsec]} & &  \multicolumn{2}{c}{Si-5~FWHM [arcsec]} & & \multicolumn{2}{c}{Qa~FWHM [arcsec]} \\
 \cline{2-3} \cline{5-6} \cline{8-9} \\
\colhead{Name} & \colhead{Source} & \colhead{PSF} & & \colhead{Source} & \colhead{PSF} & & \colhead{Source} & \colhead{PSF} 
}

\startdata
HD~38206 & 0.442 ~$\pm$~ 0.008  & 0.427 ~$\pm$~ 0.014 & & &  &   & 0.592 ~$\pm$~ 0.042 & 0.534 ~$\pm$~ 0.006 \\
HD~38678 & 0.311 ~$\pm$~ 0.001  & 0.308 ~$\pm$~ 0.001 & & &  &   & 0.605 ~$\pm$~ 0.015 & 0.536 ~$\pm$~ 0.016 \\
HD~71155 & 0.348 ~$\pm$~ 0.003  & 0.332 ~$\pm$~ 0.003 &   & & &    & 0.647 ~$\pm$~ 0.040 & 0.583 ~$\pm$~ 0.025 \\

HD~75416 &    &    & & 0.494 ~$\pm$~ 0.014 & 0.471 ~$\pm$~ 0.008 &  & 0.875 ~$\pm$~ 0.097 & 0.648 ~$\pm$~ 0.024 \\
HD~80950 & 0.372 ~$\pm$~ 0.008 & 0.412 ~$\pm$~ 0.007 &    &    & & & 0.738 ~$\pm$~ 0.125 & 0.617 ~$\pm$~ 0.015 \\
HD~115892 &    &  &  & 0.412 ~$\pm$~ 0.010 & 0.467 ~$\pm$~ 0.006 & & 0.580 ~$\pm$~ 0.016 & 0.597 ~$\pm$~ 0.012 \\
 &    &    &    &  & &  & 0.600 ~$\pm$~ 0.011 & 0.616 ~$\pm$~ 0.022 \\
HD~172555 &    &    & & 0.369 ~$\pm$~ 0.002 & 0.378 ~$\pm$~ 0.006 & & 0.591 ~$\pm$~ 0.014 & 0.559 ~$\pm$~ 0.013 \\
HD~178253 &    &    & & 0.439 ~$\pm$~ 0.003 & 0.446 ~$\pm$~ 0.008 & & 0.530 ~$\pm$~ 0.053 & 0.584 ~$\pm$~ 0.019 \\
HD~181296 &    &    & & 0.384 ~$\pm$~ 0.002 & 0.380 ~$\pm$~ 0.006 & & 0.584 ~$\pm$~ 0.192 & 0.519 ~$\pm$~ 0.012 \\
HD~181869 & 0.378 ~$\pm$~ 0.003 & 0.354 ~$\pm$~ 0.011 & &    &    &  & 0.506 ~$\pm$~ 0.024 & 0.613 ~$\pm$~ 0.037 \\
\enddata

\end{deluxetable}

\section{Source Measurements}

\subsection{Statistical Significance of IR Excesses}\label{sec:phot_estimates}

The photospheric flux densities at 10--12~$\mu$m and 18~$\mu$m were estimated by extrapolating the $2MASS$ K-band (2.2~$\mu$m) flux densities \citep{cut03} to 10~$\mu$m.  The flux density was assumed to vary as $\nu^{1.88}$ over this wavelength range, as is estimated by \citet{kur79} to be appropriate for an A0 star (e.g., Jura et al. 1998\nocite{jur98}).  Beyond 10~$\mu$m, we assumed a Rayleigh-Jeans relation ($\nu^{2}$) for the photosphere.    The photospheric flux density estimate and the corresponding excess flux density estimate are given for each source in Tables \ref{tab:excessm} and \ref{tab:excesst}.  As discussed in \S\ref{sec:data}, photometric uncertainties of 10\% and 15\% are assumed for the 10- and 18-$\mu$m windows, respectively.  We have compared our flux density estimates for the photosphere with those determined via Kurucz models for several sources (e.g., Smith et al. 2008), and the difference is less than 4\%, which is well below the photometric uncertainty.

The photometric measurements, when combined with estimates of the photospheric contribution, permit assessment of the level of excess emission attributable to dust.  For some of the sources that are known to have 24-$\mu$m excess emission, we do not detect statistically significant excess emission at 10- and/or 18-$\mu$m.  We comment on these cases specifically later in this section.  We further confirm that the excess emission (when present) is spatially coincident with the star and does not originate from a background object.  These sources were chosen based on space-based observations of their infrared excess, and the lower resolution of those images (due to the $\sim$10x smaller primary mirror) is more prone to confusion within the beam.    In \S\ref{sec:consistency}, we consider whether the implications for the location of the dust from photometric measurements and measurements of spatial extent present a consistent picture for each of the sources.

The following definitions apply to our characterization of the measured IR excesses:

\textbullet~\textbf{total uncertainty}: The total uncertainty is the quadratic addition of the measurement uncertainty (given in Table \ref{tab:fluxmeas}) and the photometric uncertainty for a calibrated image.  This value is referred to as $\sigma_{phot}$.

\textbullet~\textbf{no detected excess}: A source with no detected excess is defined as having an IR excess of less than two times the total uncertainty of the flux density measurement.

\textbullet~\textbf{marginal excess}: A source with marginal excess is defined as having an IR excess of two to three times the total uncertainty of the flux density measurement.

\textbullet~\textbf{significant excess}: A source with significant excess is defined as having an IR excess greater than or equal to three times the total uncertainty of the flux density measurement.

For nine (HD~56537, HD~83808, HD~95418, HD~102647, HD~115892, HD~139006, HD~161868, HD~178253, and HD~181869) of our 17 debris disk sources, we do not detect statistically significant excess emission in either of the bandpasses used in this work.  Two sources (HD~38206 and HD~80950) have an excess detected in only one bandpass.  The detections of excess emission are summarized in Table~\ref{tab:excessdetections}.  As discussed in \S\ref{sec:sample}, the debris disk candidates were chosen based on excess emission observed at 24 or 25 $\mu$m.  
%%We estimated the excess emission at 12 and 18~$\mu$m from the value of the $\sim$25-$\mu$m flux density and the assumption that the mid-IR colors are like those of the well studied debris disk of $\beta$~Pic \citep{tel05}.  
In the following sections, we report the detection of spatial extension for several sources (HD~56537, HD~95418, HD~139006, and HD~161868) that do not have statistically significant excess IR emission.  These results are not necessarily inconsistent, due to the uncertainties in both quantities.

\subsection{Source Extent}
 \label{sec:extent}

Only a handful of circumstellar debris disks have been spatially resolved at a level that permits examination of detailed structure. However, it is important to keep in mind that valuable information is still obtained when only the scale size is determined. A ``disk'' can consist of several components that reflect the complex relationships among the dust population, the dust parent bodies, and the planetary system, with the proposed (but still unresolved) asteroid-belt and Kuiper-belt dust zones in the triple-planet system HD~8799 being a spectacular example \citep{mar08,che09,rei09}. Establishing the existence of any of these subsystems by constraining the emitting-region size permits assessment of broader system properties, as illustrated in our analysis of Zeta~Lep \citet{moe07a} where the resolved dust may well betoken an asteroid belt and, consequently, planets.

\begin{figure}
\includegraphics[width=\columnwidth]{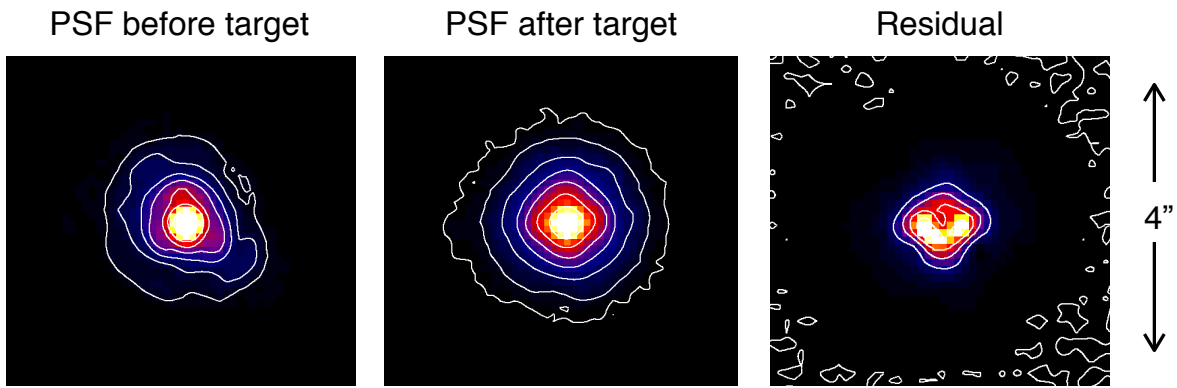}
\caption[Examples of PSF asymmetry.]{PSF observations taken at 11.2~$\mu$m with Michelle before and after HD~141569, shown to illustrate typical asymmetries encountered.  Contours on the two PSF images are drawn logarithmically to show the structure, with the lowest contour drawn at 0.3\% of the peak surface brightness.  Contours on the residual image are drawn at linear intervals of the normalized peak surface brightness of the two images, every 5\% from 0--20\%.  Note that the first PSF (left) shows slight N-S elongation in the core and a partial trefoil pattern in the wings.  The second PSF appears slightly cross-shaped in the core, resulting in contours at the center that are more square-shaped.  The residual difference between these two peak-normalized PSFs is up to 20\% of the original peak flux within the central arcsecond.
\label{fig:PSF}}
\end{figure}

To check for the presence of an extended disk source in our targets, we observed a source that is known to be not extended, a PSF star, in close temporal proximity to the target observation, as described in \S\ref{sec:data}.  In most cases, there was no obvious 2-D structure to the disk source, and the PSF references often showed asymmetric features.  Examples of asymmetric PSFs, whose causes may be associated with chopping and nodding, are shown in Fig.~\ref{fig:PSF}.  A key measure of scale size is the full-width at half-maximum (FWHM) intensity of the emitting source. Especially for fainter sources, the FWHM may be the only available measure of source size. Given the small source sizes anticipated in this study, we have focused exclusively on the use of the FWHM to characterize their extent, while remaining open to the possibility that more extended lower-level emission might be present. We used as our primary metric the FWHM measurement from a 1-D Moffat profile fitted to the  azimuthal average of each source with an IRAF routine  (i.e., we do not simply measure directly the FWHM, which, due to noise on the profile, would be a much less accurate measurement).  Moffat profile fitting has been used frequently in mid-IR image analysis (e.g., Radomski et al. 2008\nocite{rad08}), because we find that the true profile width at the half-maximum level is better approximated by a Moffat profile than by a Gaussian, regardless of the overall goodness-of-fit as evaluated by, e.g., chi-squared analysis.  We have again verified this hypothesis for several sources in the dataset presented in this work, including the known resolved sources HD~38678 and HD~141569, and the FWHM of the Moffat profile fit is closer to the true FWHM value in $>$90\% of the images measured.  This is likewise the finding of a report on PSF image quality in the mid-IR at Gemini South (which is summarized in Li, Telesco \& Varosi 2010).  We acknowledge that, in practice, one could choose a different metric such as the full-width at quarter-maximum, for which a Moffat profile fit may not be the optimal choice.

We then compared the PSF and the disk candidate FWHM values to make an assessment of the extent of the emission.  However, the science targets are typically at least an order of magnitude fainter than the reference star and must be observed for correspondingly longer integration times to achieve similar signal-to-noise ratios.  Pupil rotation, incorrect guiding correction, and changes in the quality of seeing during long observations on a science target can result in a final image degraded by lower-frequency components that are not accurately represented in the PSF determined from generally shorter integration times.  Minor variations may sometimes average out, but in the majority of cases these effects broaden the source profile in a final stacked image (see also Li, Telesco \& Varosi 2010).  
%% Aspects of this issue are explored in more detail for the Gemini mid-IR PSF by Li, Telesco \& Varosi (in press).

To estimate more robustly the profile widths and thereby assess spatial extent, we examined the FWHM of the PSF star and the target sources throughout their integration sequences.   Usually, the smallest unit of integration time was that corresponding to the so-called saveset image\footnote{A saveset is a stack of chopped images, on- and off-source, taken within one telescope nod position.  Each saveset corresponds to $\sim$10~s of integration time, and there are typically three savesets in one nod position before the telescope switches to the opposite nod position.}, but in some cases it was necessary to bin two or more savesets to achieve a signal-to-noise ratio high enough to perform a Moffat fit to the source profile.    In this way, we were able to determine a mean FWHM and corresponding uncertainty from the set of subdivided images for both the PSF and the debris disk target.  Using all of the data points, and taking into account the number of points in the set, we determined the standard deviation of the mean (the standard deviation divided by $\sqrt[]{n}$ number of measurements) for each set of FWHM measurements, and we adopt this value as the uncertainty.  The standard deviation (and standard deviation of the mean) is valid for a Gaussian distribution of independent values extracted from a population that does not vary with time, but the assumption of a stationary Gaussian distribution of values for our data sets was not always valid.  The variation in FWHMs for each of our data sets in this work sometimes revealed outlying data points, or an obvious deterioration or improvement in image quality due to factors like seeing.  However, such time-variant changes are not observed for the majority of sources in this work, and in particular we do not observe obvious changes in image quality in the data sets of sources that we claim to have spatially resolved.  Likewise, in an effort to make our data quality transparent, plots of the FWHM measurements for each source are shown in the appendix.

We have also applied a Student's $t$-test (assuming unknown and unequal variances) to compare the FWHM values of the source and PSF for each dataset to better assess whether the source and PSF data are drawn from the same image quality distribution.  We quote these results for individual sources (\S\ref{sec:consistency}) when relevant, which occurs in two cases: (1) the PSF profile data are systematically broader than the source profile data, and the $t$-test confirms that the two sets are not drawn from the same population, in which case the data are rejected, or (2) the source profile data are significantly broader than the PSF profile data, and the $t$-test confirms that the two sets are not drawn from the same population, in which case we consider the source spatially resolved.  

The following terms are defined, for use in characterization of the measured extent of the sources:

\textbullet~\textbf{combined standard deviation of the mean}: The combined standard deviation of the mean is the combination of the standard deviation of the mean of the PSF FWHM measurements and the standard deviation of the mean of the source FWHM measurements, given by

\begin{equation}
\label{eq:errorcombine}
\sigma_{ext} = \sqrt[]{\sigma_{PSF}^2 + \sigma_{source}^2}.
\end{equation}
This value is referred to as $\sigma_{ext}$ in the following discussion of extension measurements.

\textbullet~\textbf{unresolved}: An unresolved source is defined as having an average source FWHM value that is greater than its corresponding PSF FWHM by less than three times the combined standard deviation of the mean of those two measurements:

\begin{equation}
\label{eq:unresolved}
FWHM_{source} - FWHM_{PSF} < 3 \sigma_{ext}
\end{equation}

%\textbullet~\textbf{marginally resolved}: A marginally resolved source is defined as having an average source FWHM value that is greater than its corresponding PSF FWHM by between two and three times the combined standard deviation of the mean of those two measurements:

%\begin{equation}
%\label{eq:margresolved}
%2 \sigma_{ext} < \big( FWHM_{source} - FWHM_{PSF} \big) < 3 \sigma_{ext}
%\end{equation}

\textbullet~\textbf{resolved}: A resolved source is defined as having an average source FWHM value that is greater than its corresponding PSF FWHM by three or more times the combined standard deviation of the mean of those two measurements:

\begin{equation}
\label{eq:resolved}
3 \sigma_{ext} \leq FWHM_{source} - FWHM_{PSF} 
\end{equation}

\begin{figure}
\begin{centering}
\includegraphics[width=0.65\columnwidth]{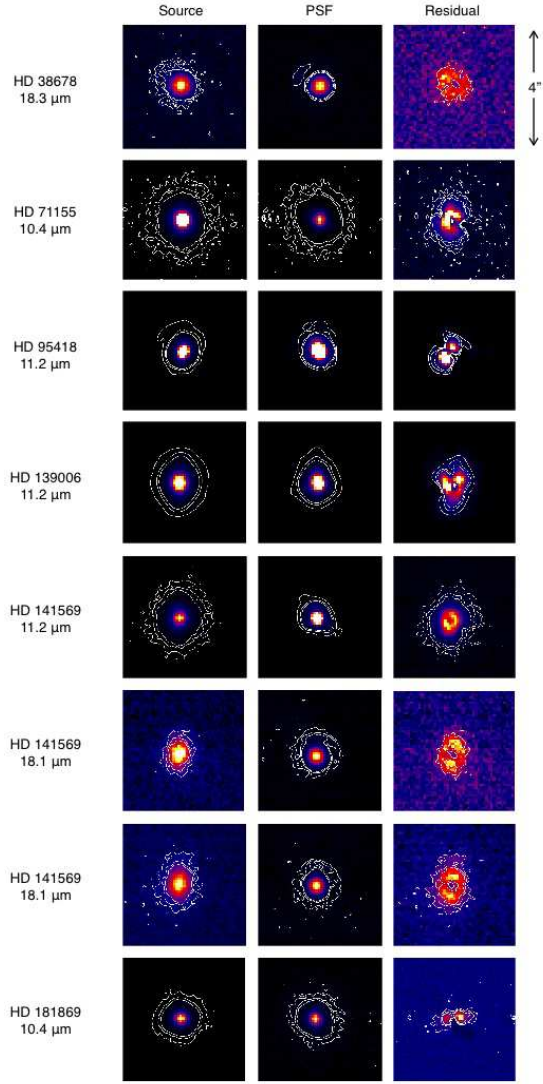}
\caption[Resolved sources.]{Images of the sources resolved in this work, the corresponding PSF stars, and the residual from peak-normalized subtraction.  Contours are drawn at 3, 6, and 9$\sigma_{bkd}$ in all images.
\label{fig:resolvedimages}}
\end{centering}
\end{figure}

Based on these FWHM measurements (Tables \ref{tab:FWHMM} and \ref{tab:FWHMT}), several sources appear to be extended.  The statistical significance of these extended sources has been assessed by breaking up the full integration time into individual images, as described in \S\ref{sec:extent}.  A list of the sources that appear to be resolved (based on FWHM measurements of profile fits to the data) and their sizes is given in Table \ref{tab:extent}.  The images of the resolved sources, their corresponding PSFs, and the residuals from peak-normalized subtraction of the PSFs are shown for reference in Fig~\ref{fig:resolvedimages}.

The profile widths of Moffat fits to the disk sources, their PSF stars, and the associated uncertainties (one standard deviation of the mean) are listed in Tables \ref{tab:FWHMM} and \ref{tab:FWHMT}.  The statistical significance of the difference between target and PSF profile width is given in Table \ref{tab:extent}.  Of the 17 debris disk candidates (i.e., excluding the two sources mentioned in \S\ref{sec:sample}) that we imaged,
five sources near 10~$\mu$m  ($\lambda_c$ = 10.7, 11.2, or 11.7~$\mu$m) and two sources near 20~$\mu$m ($\lambda_c$ = 18.1 or 18.3~$\mu$m) had source FWHM values bigger than the PSF FWHM by more than three times the combined standard deviation of the mean for the two measurements.  These sources are discussed further in \S\ref{sec:consistency}.

\begin{deluxetable}{l ccc c ccc }
\newcommand\T{\rule{0pt}{3.4ex}}
\newcommand\B{\rule[-1.8ex]{0pt}{0pt}}

\tablecaption{Sizes of Extended Sources\label{tab:extent}}

\tablewidth{0pc}
\tablehead{
\colhead{} & 
\multicolumn{2}{c}{10.4~$\mu$m} & &
\multicolumn{2}{c}{11.2~$\mu$m}  \\
\cline{2-3} \cline{5-6} 
\colhead{} & 
\colhead{$\Delta FWHM\tablenotemark{a}$} & 
\colhead{r} & &
\colhead{$\Delta FWHM\tablenotemark{a}$} & 
\colhead{r}  \\
\colhead{HD} & 
\colhead{[arcsec]} & 
\colhead{[AU]} & &
\colhead{[arcsec]} & 
\colhead{[AU]} & 
}

\startdata
71155 & 0.10 $\pm$ 0.01 & 1.96 $\pm$ 0.02 &   & ... &  ...  \\ 
95418 & ... & ... &    & 0.09 $\pm$ 0.01 & 1.09 $\pm$ 0.01  \\ 
139006 & ... & ... &   & 0.20 $\pm$ 0.01 & 2.31 $\pm$ 0.04 \\ 
141569 & ... & ... &   & 0.22 $\pm$ 0.02 & 10.89 $\pm$ 0.48  \\ 
%161868 & ... & ... &   & 0.11 $\pm$ 0.01 & 1.66 $\pm$ 0.03  \\ 
181869 & 0.16 $\pm$ 0.01 & 4.12 $\pm$ 0.11  &   & ... & ...  \\ 
\\

\colhead{} &
\multicolumn{2}{c}{18.1~$\mu$m} & &
\multicolumn{2}{c}{18.3~$\mu$m}  \\
\cline{2-3} \cline{5-6} 
\colhead{} & 
\colhead{$\Delta FWHM\tablenotemark{a}$} & 
\colhead{r} & &
\colhead{$\Delta FWHM\tablenotemark{a}$} & 
\colhead{r}  \\
\colhead{HD} & 
\colhead{[arcsec]} & 
\colhead{[AU]} & &
\colhead{[arcsec] \B}  & 
\colhead{[AU]} & \\
\hline
38678 \T & ... &  ... &   & 0.28 $\pm$ 0.02 &  3.01 $\pm$ 0.24  \\ 
% 56537 & 0.25 $\pm$ 0.03 &  3.67 $\pm$ 0.16 &    &  ... &  ...  \\ 
% 75416 & ... & ... &   & 0.59 $\pm$ 0.06 & 28.6 $\pm$ 2.1 \\ 
139006 & 0.12 $\pm$ 0.06 & 1.36 $\pm$ 0.09 &   & ... & ... \\ 
141569 & 0.60 $\pm$ 0.03 & 29.7 $\pm$ 1.1 &   & ... & ...  \\ 
             & 0.63 $\pm$ 0.07 & 31.1 $\pm$ 2.5  &   & ... & ...  \\ 

\enddata \\

\tablenotetext{a}{This difference in FWHM is calculated by subtracting the PSF FWHM from the source FWHM in quadrature. 
	}

\end{deluxetable}

\section{Expected detectability based on comparison to archetypes}

Here we consider a simple assessment of how many sources from our sample we would expect to spatially resolve if the source morphologies were similar to those of certain archetypal disks that have already been resolved with mid-IR imaging.  We use two archetypes for comparison: (1) a possible asteroid belt analog, $\zeta$~Lep, and (2) a possible Kuiper Belt analog, $\beta$~Pic.  The $\zeta$~Lep emission is very compact, and the evidence for a disk in this type of a source would be found in
the FWHM.  In contrast, the central disk of $\beta$~Pic is very extended but relatively faint and for the most part not
evident in a measurement of the FWHM, but rather in the wings of the profile.

\subsection{$\zeta$ Lep Analog}

$\zeta$~Lep was resolved in 18.3-$\mu$m images, and the measured spatial extent implied a dust disk radius of 3~AU \citep{moe07a}.  The disk radius was inferred from a quadratic subtraction of the PSF FWHM from the azimuthally averaged source FWHM:  ${FWHM}_{source}^2 - {FWHM}_{PSF}^2 = {D}_{disk}^2$.  (Such a relationship is strictly true for Gaussian functions, and another example of combination in quadrature is given in Eq.~\ref{eq:errorcombine}.)  Therefore, we consider whether each of the sources in our sample (besides $\zeta$~Lep) would be spatially resolved if it were described by a point source (the size of which is dictated by the corresponding PSF) and a dust disk of radius 3 AU with the same brightness relative to its host star as that of the $\zeta$~Lep system.  We calculate the FWHM for each source as the quadratic addition of its corresponding PSF FWHM and a disk with a 3-AU radius: ${FWHM}_{PSF}^2 + D_{6~AU}^2 = {FWHM}_{source}^2$.  The point source considered here may correspond to stellar emission and/or emission from dust from an unresolved region near the star.

We consider a source to be spatially resolved if the value $FWHM_{source}$ (convolved PSF and 3-AU-radius-disk) is greater than its corresponding value for $FWHM_{PSF}$ by more than 3$\sigma_{ext}$, where $\sigma_{ext}$ is the total uncertainty associated with the FWHM measurements (discussed further in \S\ref{sec:extent}).  Based on the PSF observations associated with each source and the assumption that they have a morphology like that of $\zeta$~Lep, we estimate that eleven sources should have been resolved in images near 10~$\mu$m and five sources should have been resolved in images near 18~$\mu$m.  In reality, five sources were resolved in 10-$\mu$m images, and two sources were resolved  in 18-$\mu$m images (results discussed further in \S\ref{sec:extent}).  HD~141569 is in both of these sets, and it is known from prior observations that this dust disk is more extended (r$>$30~AU) and is not comparable in nature to $\zeta$~Lep.  Removing HD~141569 from the set of resolved sources leaves three sources resolved in 10-$\mu$m images and one source resolved in 18-$\mu$m images.  The fact that only approximately half of the projected number of resolved Zeta-Lep-like sources were actually resolved suggests that not all of the sources are comparable in size to the  asteroid-belt-like analog associated with $\zeta$~Lep.  However, it is also possible that some sources have a disk size comparable to that of $\zeta$~Lep but remain undetected due to lower disk brightness.  This is a reasonable possibility, because $\zeta$~Lep has a higher fractional IR luminosity than most of the sources in our sample.

\subsection{$\beta$ Pic Analog}

We also assess the number of sources we would expect to resolve if the disk morphology of each target were comparable to that of the well known disk $\beta$~Pic.  For each target, we check whether extended emission (r$\sim$100~AU) like that seen around $\beta$~Pic would be significantly detected above the background noise level in our actual source images.  To be clear, this is not spatial extent sought at the FWHM level, but emission farther out in the brightness profile that is significantly above the level of the background.  We acknowledge that edge-on disks like $\beta$~Pic may be more detectable than face-on disks because of the higher line-of-sight column densities, and so we consider the disk emission levels for both an edge-on and a face-on case.  We used unaltered mid-IR images of $\beta$~Pic \citep{tel05} for the edge-on case, since it is nearly edge-on already.  For the face-on case, we constructed simple model images of a face-on $\beta$~Pic by integrating the measured flux density as a function of radius and then redistributing it azimuthally.  By generating this face-on model, we assume that the disk is optically thin and its MIR emission is azimuthally symmetric.  In images of $\beta$~Pic taken from the ground with subarcsecond resolution (e.g., Telesco et al. 2005), we see that the assumption of azimuthal symmetry is a simplification of the disk structure but is reasonable for our purposes here.

We generated profile cuts of the edge-on and face-on images to compare the flux density levels to the background noise measured in our target images.  Profile cuts were made by sampling the line of pixels along the major axis of the disk for the edge-on case and along an axis bisecting the disk in the face-on case.  To account for the different central star brightness and accordingly different disk brightness of our target sources, we scaled the brightness of the $\beta$~Pic profiles according to the stellar flux density of each of the sources.  We expect that the 2-$\mu$m flux density is predominantly from the photosphere, and we therefore used the $2MASS$ measurement for each target as a metric for the stellar brightness and, accordingly, its dust heating ability.  We scaled the profiles for each of the sources by the ratio of the target source $2MASS$ flux density to that of $\beta$~Pic, $F_{target}(2~\mu m)/F_{\beta~Pic}(2~\mu m)$.  The width of the extended emission profiles was also scaled for each source according to the source distance.

We characterize a $\beta$-Pic-type source as one that we could have detected if the extended emission of the test profile is (1) above the 5-$\sigma_{bkd}$ level (five times the per-pixel background noise measured in the actual source images) and (2) beyond the first Airy null, since some simulated profiles might have been bright enough for detection but were not spatially extended beyond the Airy disk.   Five-$\sigma$ levels are used to assess extension because the mid-IR background is noisy enough that 3-$\sigma$ ``blobs'' that are not associated with disk emission are relatively common.  If all of the sources in our sample had the same morphology and fractional IR luminosity as $\beta$~Pic, we would expect the following number of detections: for an edge-on orientation, eleven at 10~$\mu$m and six at 18~$\mu$m, and for a face-on orientation, five at 10~$\mu$m and none at 18~$\mu$m.  (We note that the disk of $\beta$~Pic itself is not detected in a face-on orientation in this test, keeping in mind that the considerable observed levels of optically thin emission have been redistributed azimuthally for the face-on model, resulting in much lower surface brightnesses than in its true nearly edge-on orientation.)
In fact, our images reveal that only one source, HD~141569, shows significant extended emission (in both bandpasses).   However, a random distribution of disks would only have approximately 10\% of disks with inclinations within ten degrees of edge-on; therefore, the number of detections that we expect in a realistic distribution of disk inclinations should be $\sim$10\% of our predicted number of edge-on detections, $\sim$1 at 10$\mu$m and $<$1 at 18~$\mu$m.  Therefore, perhaps surprisingly, our observed results are consistent with a population composed entirely of $\beta$-Pic-type disks with a random distribution of orientations.  However, this statement is weakened by the fact that our sample contains a relatively small number of sources.

Our general conclusion based on comparison to $\zeta$~Lep and $\beta$~Pic archetypes is that our results are generally consistent with the expected detection rate of several more asteroid belt analogs like $\zeta$~Lep but only one (HD~141569) Kuiper Belt analog like $\beta$~Pic.  We examine this issue more below.

\section{Consistency of Spatial and Photometric Measurements}
\label{sec:consistency}

Here we assess the consistency of the observed color temperature of the excess emission with the temperature of dust in radiative equilibrium at the distance implied by the observed spatial disk extent.  The color temperature is an upper limit to the true temperature because it is the unique solution to the equation

\begin{equation}
\label{eq:colortemp}
\frac{F_{{\nu}_1}}{F_{{\nu}_2}} = \frac{Q_{{\nu}_1}}{Q_{{\nu}_2}} \frac{B_{{\nu}_1}(T)}{B_{{\nu}_2}(T)}
\end{equation}

where $B_{{\nu}_1}(T)/B_{{\nu}_2}(T)$ is the ratio of two points on the Planck function for a temperature $T$, and $Q_{{\nu}_1}/Q_{{\nu}_2}$, the ratio of the two emission efficiencies, is unity.  If the particles behave as blackbodies, then the ratio $Q_{{\nu}_1}/Q_{{\nu}_2}$ is unity.  For particles comparable in size or smaller than the emission wavelength, the emission efficiency is sometimes described as $Q_{em}\propto\nu^n$, with n=1--2, and if $\nu_1 > \nu_2$, then $Q_{{\nu}_1}/Q_{{\nu}_2}$ will be greater than unity.  In this non-blackbody case, the ratio of the two points on the Planck function $B_{{\nu}_1}(T)/B_{{\nu}_2}(T)$ would have to be lower, and thus originate from a lower-temperature source, in order to produce the same observed flux density ratios.  Real dust particles are generally not blackbodies, and the computed color temperature is therefore an overestimate of the true physical temperature.

While the relationship between a diskÕs observed color temperature and a ``true'' dust temperature depends on the distributions of particle sizes and locations, numerous examples suggest that the observed global color temperature of a disk can give an indication, albeit a rough one, of the typical distances of the mid-infrared-emitting particles from the star, and therefore of the disk size.
For example, the 100-K mid-IR color temperature of $\beta$~Pic implies that dust with blackbody behavior would be located at $\sim$25~AU, which is within the bounds of the $\sim$100~AU radial extent of the mid-IR disk emission \citep{tel05}; likewise, the 327-K mid-IR color temperature of $\zeta$~Lep implies a dust distance of 2.9~AU, which is consistent with the 3-AU radial extent determined from 18-$\mu$m images \citep{moe07a}.

\begin{figure}
\begin{centering}
\includegraphics[width=\columnwidth]{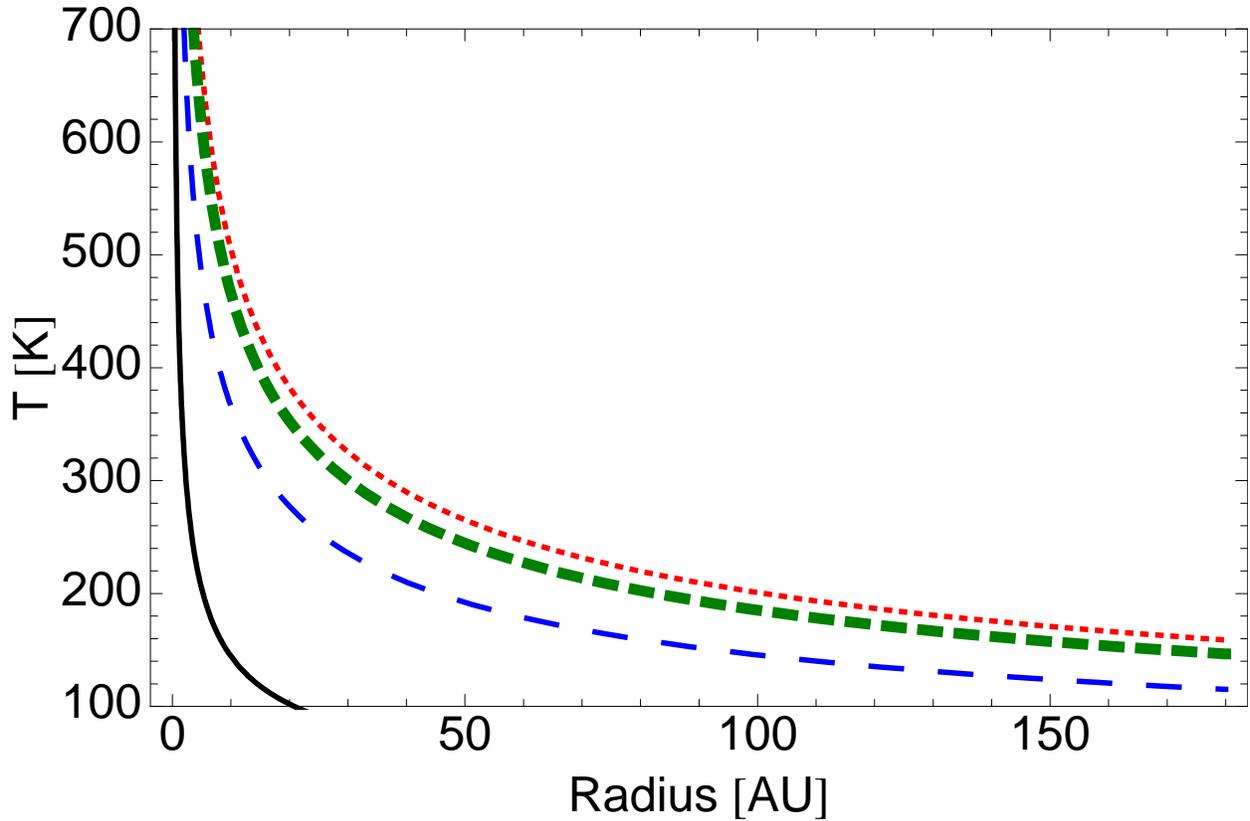}
\caption[Particle temp reference]{The expected blackbody temperature as a function of distance from a $\sim$7-L$_{\odot}$ main-sequence star (solid line), and
the predicted temperatures for efficiently absorbing but inefficiently emitting particles (as discussed in Backman \& Paresce 1993) with characteristic grain sizes of (from left to right) 0.25~$\mu$m (long dash), 0.075~$\mu$m (medium dash), and 0.05~$\mu$m (short dash).
\label{fig:tempref}}
\end{centering}
\end{figure}

Disk extent ($r_{AU}$) is estimated by quadratic subtraction of the PSF FWHM from the source FWHM.  The resulting estimate for the dust temperature $T_d$, for blackbody particles at that distance from the star, is given by

\begin{equation}
\label{eq:blackbody}
T_d = 278~L_{\star}^{\frac{1}{4}}~r_{AU}^{-\frac{1}{2}}
\end{equation}
where the stellar luminosity $L_{\star}$ is in units of $L_{\odot}$, solar luminosity, $r_{AU}$ is the radius of the dust annulus in AU, and the equilibrium temperature at 1 AU (Earth) is 278~K.  Uncertainties for this temperature estimate are calculated by propagating the uncertainty in the disk extent ($r_{AU}$) through Equation~\ref{eq:blackbody}.  This temperature estimate from Equation~\ref{eq:blackbody} is a lower limit to the true temperature, because we have assumed that the dust particles are blackbodies.  In reality, as noted above, ``small'' particles are heated to higher temperatures than blackbodies at the same distance from the star.  
%In the cases of unresolved sources where we can only assign an upper limit to the disk radius, the dust temperature computed with Equation~\ref{eq:blackbody} offers only a lower limit to the temperature, because the dust located at radii less than the limit would be accordingly hotter.  
For example, we have plotted in Figure~\ref{fig:tempref} the relationship between temperature and distance from a $\sim$7-L$_{\odot}$ star (representative of an A-type main-sequence star) for both blackbody-type particles and less efficient emitters with characteristic sizes of 0.05, 0.075, and 0.25~$\mu$m.  The temperatures for these inefficient emitters were calculated based on equations from \citet{bac93}, which estimate particle temperature as a function of distance based on assumptions regarding particle size and composition, and thereby radiation efficiency.  The case adopted for our purely demonstrative calculations is that of a particle which absorbs efficiently but emits inefficently, such as graphite or amorphous silicate.  It can be seen in this plot that for a given observed dust temperature, the implied distance from the star depends significantly on the particle properties, especially at the lowest temperatures, and this should be kept in mind when considering our calculations in the following sections.

In cases where we detect excess emission that is not spatially extended, we estimate a temperature with Equation~\ref{eq:blackbody} and the 2-$\sigma_{ext}$ limit for the observation.  That is, we assume that any extension (source FWHM minus the PSF FWHM) that is less than 2$\sigma_{ext}$ could escape detection.  Thus, we estimate the upper limit for disk extent as

\begin{equation}
\label{eq:siglimit}
r_{limit} = \frac{1}{2} ~\sqrt[]{(FWHM_{PSF} + 2\sigma_{ext})^2 - FWHM_{PSF}^2}.
\end{equation}
This estimate also yields a lower limit to the true dust temperature, because the temperature will increase if (1) the dust is any closer to the star, or (2) the particles are small enough that they do not behave like blackbody emitters.  The radius limits calculated with this method are summarized in Table~\ref{tab:unresolvedradii}.

\begin{table}[]

	\caption{Dust Radius Limit Estimates (in AU) for Unresolved Sources with IR Excess}\label{tab:unresolvedradii}
	\begin{tabular}{l c c c c}
	\hline
HD & Si-5 (11.7~$\mu$m)  & Qa$^a$ (18.3~$\mu$m) \\
 \hline

38206  & ... &  10.8 \\
%56537  & 2.2 & ... & ... & ... \\
71155  &  ... &  6.2 \\
75416  & 8.6 &  26.5 \\
80950  &  ... &  8.1 \\
%161868  & ... & ... & 3.3 & ... \\
172555  & 1.4 &  3.1 \\
178253  &  2.4 &  7.5 \\
181296  &  2.3 &  17.6 \\
181869  &  ... &  8.9 \\
 \hline
	\end{tabular}
	\\
	\small{Notes-- $^a$~T-ReCS, Gemini South.  All limits were estimated with Equation~\ref{eq:siglimit}.}
	
\end{table}

\subsection{Sources Unresolved at Both Wavelengths}

\textbf{HD 38206}--
There was no excess emission detected at 10.4~$\mu$m.  We estimated a temperature for the unresolved 18.3-$\mu$m-emitting dust by calculating the blackbody temperature at the orbital distance corresponding to the 2-$\sigma_{ext}$ limit of the source FWHM at 18.3~$\mu$m, as described by Equation~\ref{eq:siglimit}.  This temperature limit is 201~K. 

\textbf{HD 56537}--
HD~56537 was not resolved at either wavelength, and we did not detect significant excess emission.  

\textbf{HD 75416}--
The equilibrium temperature at the orbital distance corresponding to the 2-$\sigma_{ext}$ limits of the source FWHM at 11.7~$\mu$m and 18.3~$\mu$m are 307~K and 175~K, respectively.  To estimate the color temperature of 653~$\pm$~272~K, we assumed that the particles behave as efficient absorbers and inefficient emitters, because the color could not be fitted by a single-temperature perfect blackbody.  Since the color temperature is an upper limit to the true dust temperature, the temperature estimates based on spatial extension measurements are consistent.

\textbf{HD 80950}--
For the purpose of assessing spatial resolution via FWHM measurements, we rejected the 10.4-$\mu$m data, because the PSF was systematically broader than the disk candidate.  This was quantitatively confirmed by a Student's $t$-test with $p$-value of 0.0001, or greater than 99.9\% certainty that the datasets are drawn from distinct image quality distributions.

In 10.4-$\mu$m images of HD~80950, we detected two sources separated by 1.6''.  The flux density of the primary is 170~$\pm$~18~mJy, and the flux density of the secondary is 8~$\pm$~1~mJy (uncertainties include both photometric and background uncertainty).  We did not detect any excess emission associated with the primary at 10.4~$\mu$m.  If both sources are measured within the same aperture, their measured flux density is 190~$\pm$~23~mJy, where we see a greater contribution to the background noise due to the larger aperture.  Our flux density measurements are consistent with the $IRAS$ 12-$\mu$m flux density (Faint Source Catalog) of 211~$\pm$~25~mJy, and $IRAS$ would not have distinguished the two sources due to its large beam size.  Indeed, no record of another source within the T-ReCS field is found in the $IRAS$ or $2MASS$ catalogs.  It is therefore possible that the IR excess previously thought to be associated with HD~80950 arises from this apparent companion object.  We intend to characterize this object with further mid-IR and near-IR photometry and/or spectroscopy, but we note that the frequency of unassociated background objects encountered in the mid-IR is low.
If the unresolved 18.3-$\mu$m emission originates with dust at an orbital distance corresponding to the 2-$\sigma_{ext}$ limit of the source FWHM, then the implied blackbody temperature is at least 244~K.

\textbf{HD 83808}--
While no excess emission was detected in either bandpass, the uncertainties for the observed excess are large (Table~\ref{tab:excessm}).  $Spitzer$ detected excess emission at 24~$\mu$m, and there are therefore two possibilities for why we do not detect any excess at 12 or 18~$\mu$m: (1) the dust emission is diffuse enough that we do not have sufficient sensitivity to detect it above the background noise, or (2) the dust is too cool to emit significantly at 12 and 18~$\mu$m.   We also reject the 11.2-$\mu$m data for use in assessing spatial extent due to the PSF being systematically broader than the disk target, which is confirmed by a Student's $t$-test $p$-value of 0.008.

\textbf{HD 102647 ($\beta$~Leo)}--
We did not detect any IR excess emission associated with the source, within our error bars.  Again, the $Spitzer$ detection of excess emission indicates that there is dust present associated with the source, so there exist the same two possibilities as in the case of HD~83808: low surface brightness or cool dust temperatures.

\textbf{HD 115892}--
We rejected the 11.7-$\mu$m data for use in assessment of spatial extent, because the PSF profile is broader than the disk source profile; this is confirmed by a Student's $t$-test $p$-value of 1.5 x 10$^{-6}$.

We detected no excess emission within our error bars.  $Spitzer$ observations do show excess emission at 24~$\mu$m.  If this excess is real, then we do not detect the dust emission either because of low surface brightness or because of dust temperatures that only yield emission longward of 18~$\mu$m.  

\textbf{HD 161868}--
We did not detect excess emission in either bandpass, and the source FWHM is not greater than the PSF FWHM by more than 3$\sigma_{ext}$.  

\citet{su08} recently resolved this debris disk with $Spitzer$ MIPS images and estimated a disk radius of $\sim$520~AU from the 24~$\mu$m images and a disk radius of $\gtrsim$260~AU from the 70~$\mu$m images.  The color temperature based on the 24~$\mu$m and 70~$\mu$m photometry is 81~K, which overestimates the flux density at 28--35~$\mu$m and 55--65~$\mu$m, and the authors note that this suggests a range of dust temperatures.  Nonetheless, the 81~K blackbody fits the available SED points reasonably well.   The IRS spectrum is also shown, which indicates flux densities of $<$100~mJy at wavelengths $<$20~$\mu$m; these spectral measurements confirm that our estimated mid-IR excess levels are within expectations.  Given that the color temperature and the IRS spectrum (and MIPS images) indicate the presence of predominantly cold dust, it is not surprising that we do not find strong evidence of excess emission or resolved spatial structure in mid-IR images.  

\textbf{HD 172555}--
We detected significant IR excess in both bandpasses.  We estimated a minimum dust temperature by assuming that the dust must lie interior to the disk size corresponding to the 2-$\sigma_{ext}$ limit of the source FWHM.  The temperature estimates based on extension measurements are 407~K (based on the 11.7~$\mu$m data) and 278~K  (based on the 18.3~$\mu$m data).  The color temperature calculated from the excess emission measurements from the two bandpasses is 274~$\pm$~34~K.  Therefore, the temperature estimate based on the source size limit at 18.3~$\mu$m is consistent with the color temperature, but the temperature estimate based on the 11.7-$\mu$m source size is not.  It is possible for nonspherical particles to have temperatures lower than those expected for blackbody emission (e.g., Greenberg \& Shah 1971; Voshchinnikov \& Semenov 2000\nocite{vos00}), and this may be the case for the 11.7-$\mu$m-emitting particles in HD~172555.

\citet{wya07b} noted that the fractional IR luminosity of HD~172555 is anomalously high, or eighty-six times the expected maximum fractional IR luminosity based on steady-state evolution models.  An estimate of 6~AU is given for the disk radius, based on 24~$\mu$m and 70~$\mu$m $Spitzer$ photometry; this radius corresponds to a blackbody equilibrium temperature of $\sim$200~K.  This temperature is $\sim$75~K lower than the color temperature estimate based on mid-IR color, but it is possible that the dust present in the disk spans a range of radii, and therefore temperatures, and (at least) two different populations of dust are being sampled by the 12~$\mu$m/18~$\mu$m color and the 24~$\mu$m/70~$\mu$m color.  

\textbf{HD 178253}--
We did not detect statistically significant excess emission associated with this source.  However, if there is unresolved emission originating from dust within the region corresponding to the size of the 2-$\sigma_{ext}$ limit of the source FWHM at each of the observed bandpasses, then the emitting dust temperatures should be at least 432~K (based on the 11.7-$\mu$m FWHM) and 246~K (based on the 18.3-$\mu$m FWHM).  

\textbf{HD 181296}--
There is significant IR excess emission detected in both bandpasses.  The temperature estimates for dust within a region the size of the 2-$\sigma_{ext}$ limit of the source FWHM are 397~K (based on the 11.7-$\mu$m FWHM) and 143~K (based on the 18.3-$\mu$m FWHM).  Only the estimate based on the source FWHM at 18.3~$\mu$m is consistent with the color temperature, which is 229~$\pm$~23~K.

Since our initial observations, HD~181296 has been resolved in Q-band images taken in July 2007 at mid-IR wavelengths with longer (6.5x) integration times \citep{smi09a}.  It is likely that the initial observations did not resolve the source because the integration time was not sufficient to detect the surface brightness of the disk. The extended emission peaks detected by \citet{smi09a} total $90~\pm~5$~mJy, and if we assume that the flux is approximately equally distributed between the two peaks, then this emission is less than three times our background noise level.   \citet{smi08} discuss in greater detail the relationship between disk morphology and the predicted length of integration required to resolved a given source.  

Models of the 18.3-$\mu$m images of the source indicate that $\sim$50\% of the excess emission originates in the resolved component in an apparently edge-on disk with a 24-AU radius, which can be fit by a modified blackbody of temperature $\sim$100~K.  The rest of the excess emission resides in an unresolved component of temperature 310~K, consistent with dust at 3.9~AU.  Discrepancies in temperature estimates based on the initial and follow-up data sets exist primarily because of a 13\% lower flux density measurement at 11.7~$\mu$m (from the more recent images); however, these two measurements are still within $\sim$1$\sigma$ of each other.

\subsection{Resolved Sources
\label{sec:resolved}
}

\begin{figure}
\begin{centering}
\includegraphics[width=\columnwidth]{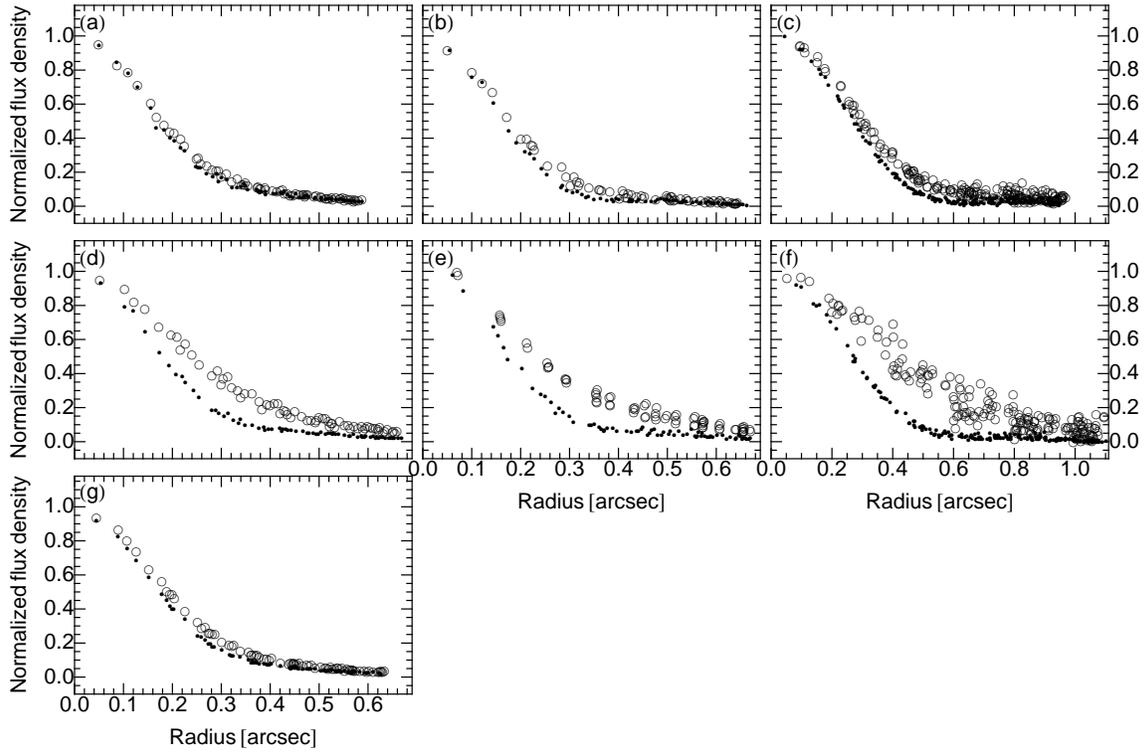}
\caption[radial profiles]{Normalized radial brightness profiles of resolved sources (open circles) and their corresponding PSF stars (filled circles).  Sources: (a)~HD~71155~(10.4~$\mu$m), (b)~HD~95418~(11.2~$\mu$m), (c)~HD~141569~(11.2~$\mu$m), (d)~HD~139006~(11.2~$\mu$m), (e)~HD~38678~(18.3~$\mu$m), (f)~HD~141569~(18.1~$\mu$m), (g)~HD~181869 (10.4~$\mu$m).
\label{fig:radplots}}
\end{centering}
\end{figure}

For reference, we have plotted the azimuthally averaged radial profiles of the resolved sources and their corresponding PSFs in Figure~\ref{fig:radplots}.  However, we note that the images used to generate these plots are only median stacks of all of the subset frames that were used to assess their spatial resolution.

\textbf{HD 141569}--
We review HD~141569 first, because its spatial extent has been detected previously in mid-IR images \citep{fis00, mar02} and this data set can therefore provide a benchmark for comparison as a consistent body of evidence that describes the source size.  

%% add comparison with values from Fisher paper.

The source FWHM appears extended in comparison to the PSF FWHM at both 11.2~$\mu$m and 18.1~$\mu$m, and the statistical significance of these extensions are 3.6$\sigma_{ext}$ (11.2$\mu$m) and 9.1$\sigma_{ext}$ and 4.5$\sigma_{ext}$ (18.1-$\mu$m, from two nights of data).  These data were not combined, because the location of the central star could not be determined accurately, and this location is necessary to register and stack images.  Student's $t$-tests on the data yield the following results: at 11.2~$\mu$m, the $p$-value is 0.005 and there is a 95\% confidence interval that the source FWHM is greater than the PSF FWHM by 0.023'' to 0.100''; at 18.1~$\mu$m, the $p$-values are 6~x~10$^{-5}$ and 6~x~10$^{-4}$, and the 95\% confidence intervals are 0.195''--0.340'' and 0.151''--0.439''.

The blackbody dust temperatures at the radii inferred from the 11.2-$\mu$m images and the 18.1-$\mu$m images are 186~$\pm$~4~K at $\sim11$~AU and 112~$\pm$~3~K at $\sim$30~AU, respectively.  (For comparison, \citet{fis00} determined disk radii of 17~AU and 34~AU at 10.8~$\mu$m and 18.2~$\mu$m, respectively.)  These temperatures are consistent with the color temperature estimate for HD~141569 of 191~$\pm$~16~K, in that the extension-implied temperatures are a lower limit and the color temperature is an upper limit.  Indeed, the models of both \citet{fis00} and \citet{mar02} for this disk suggest that the dominant particle population is composed of inefficient emitters that would be heated to temperatures above those expected for perfect blackbody emitters, which implies that our extent-implied temperatures should also be higher.

\textbf{HD 38678, $\zeta$~Lep}--
$\zeta$~Lep is unresolved at 10.4~$\mu$m but is resolved at 18.3~$\mu$m, with a significance level of 3.1$\sigma_{ext}$.  These results are discussed in detail in a prior work \citep{moe07a}.  We have further confirmed the extent with a Student's $t$-test, which yields a 95\% confidence interval that the source FWHM is greater than the PSF FWHM by 0.035''--0.114''.  The spatial extent of the 18.3-$\mu$m disk profile implies a disk radius of 3~AU, which is comparable in location to the asteroid belt in the solar system.  Prior to the spatial resolution of HD~71155 (discussed below), this was the only resolved debris disk spanning only a few AU \citep{che01,moe07a}.  The color temperature for the excess emission associated with $\zeta$~Lep is 323$^{+27}_{-30}$~K, which is consistent with the blackbody temperature of 320~K for dust grains at 3.0~AU.  The implied presence of large (r $\sim$ few microns) grains that emit like blackbodies is supported by $Spitzer$ IRS spectra taken by \citet{che06}, which show no silicate emission feature.  Relatively large grains may not cause silicate emission, as shown by \citet{prz03}.

\textbf{HD 71155}--
HD~71155 is resolved at 10.4~$\mu$m, at a significance level of 4.4$\sigma_{ext}$.   The 95\% confidence interval from a Student's $t$-test is 0.008''-0.025'', with a $p$-value of 1~x~10$^{-4}$.  The extent (as computed with the full data set) implies a disk radius of 2.0~$\pm$~0.1~AU, at which the blackbody temperature is 499~$\pm$~3~K.  The marginal excess emission for this source could not be fitted by a simple single-temperature blackbody. The excess was instead fitted with emitting particles that are efficient absorbers but inefficient emitters.  This may be the case for particles which are larger than the peak wavelength of stellar emission but smaller than the peak wavelength of particle thermal emission \citep{bac93}.  In the literature, an emission efficiency of the form $Q_{em}\propto\nu^n$, with n=1--2, is often assumed.  The color temperature for such inefficient emitters with n=1 is 487~$\pm$~129~K, which is consistent with the temperature implied by the 10.4-$\mu$m extension.  

The unresolved 18.3-$\mu$m result is consistent if this emission originates from the same location as the 10.4-$\mu$m emission; the quadratic sum of the 10.4-$\mu$m-based source size and the 18.3-$\mu$m PSF FWHM size is less than the nominal detection limit for extension, $FHWM_{PSF} + 2\sigma_{ext}$.
If this unresolved 18.3-$\mu$m-emitting dust is assumed to lie within a region corresponding to this 2-$\sigma_{ext}$ limit of the source FWHM, then its equilibrium temperature is at least 281~K, which is also consistent with the upper limit set by the color temperature.

\begin{figure}
\begin{centering}
\includegraphics[width=\columnwidth]{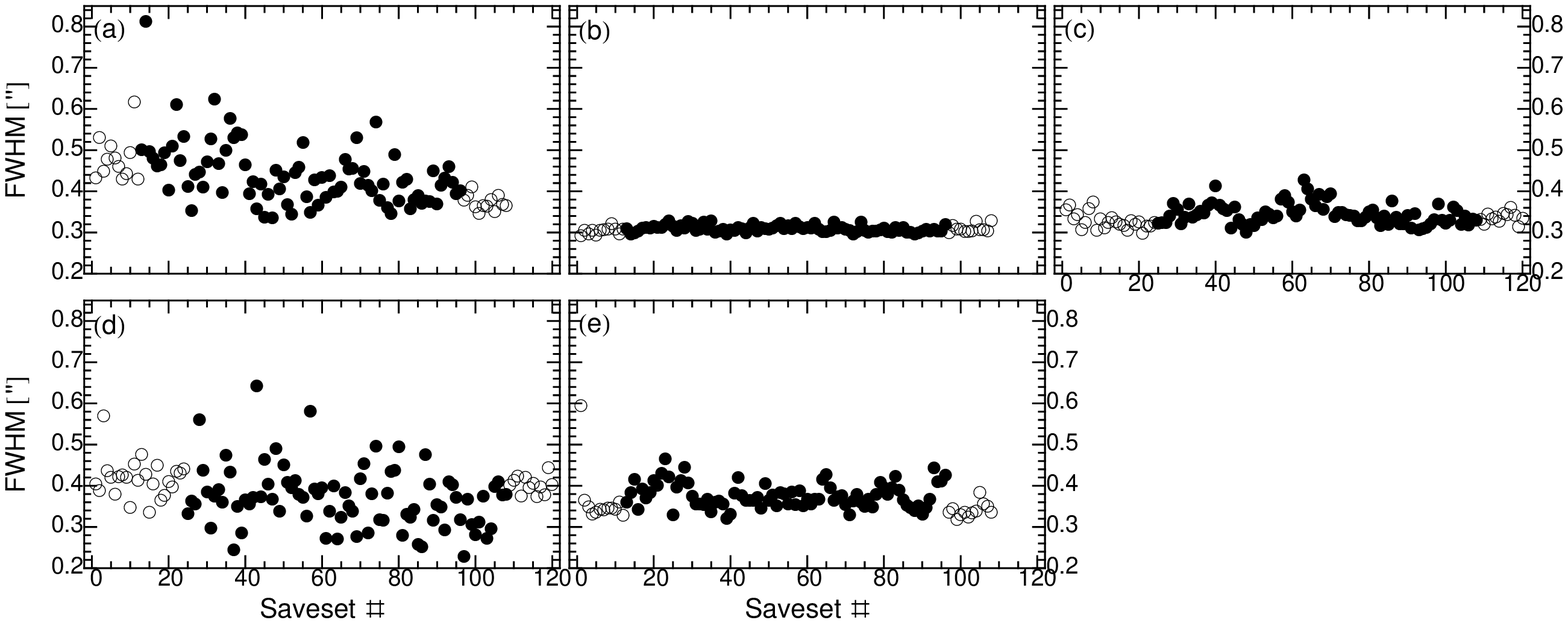}
\caption[Broadband N sets]{FWHM of profile fits to the sources at 10.4~$\mu$m, per saveset.  Open circles represent the PSF reference star, and filled circles represent the debris disk target. Sources: (a)~HD~38206, (b)~HD~38678, (c)~HD~71155, (d)~HD~80950, (e)~HD~181869.
\label{fig:Nsets}}
\end{centering}
\end{figure}

In Fig.~\ref{fig:Nsets}, we note the possibility that the image quality has systematically degraded around savesets \#40 and \#65.  We tested the impact of the data in this region by removing 25 contiguous savesets that include the peaks in FWHM and then recomputing the mean FWHM value for the source, an exercise which demonstrates the utility of measuring the source profile in every saveset rather than solely in the final stacked image.  We found that the source FWHM decreased from 0.348'' to 0.339'', but that the standard deviation likewise decreases, such that the source FWHM is still greater than the PSF FWHM at a 3.1-$\sigma_{ext}$ level.  When the $t$-test is repeated with the truncated data set, the $p$-value is 0.035, and there is a 95\% confidence level that the source FWHM is greater than the PSF FWHM by 0.001''--0.015''.  Nonetheless, this result should be confirmed with deeper imaging observations.

\textbf{HD 95418}--
The source is resolved at 11.2~$\mu$m with a statistical significance of 6.8$\sigma_{ext}$. The results of a Student's $t$-test on the data are a $p$-value of 3~x~10$^{-5}$ and a 95\% confidence interval of 0.008''--0.016''.  No excess emission associated with HD~95418 is detected in either bandpass, but the measurement of the excess emission at 11.2$\mu$m is within 2$\sigma_{phot}$ of a statistically significant excess detection, so the spatial resolution at that wavelength is consistent. The dust temperature at the orbital distance implied by this resolved result is 764~$\pm$~2~K.

\begin{figure}
\includegraphics[width=\columnwidth]{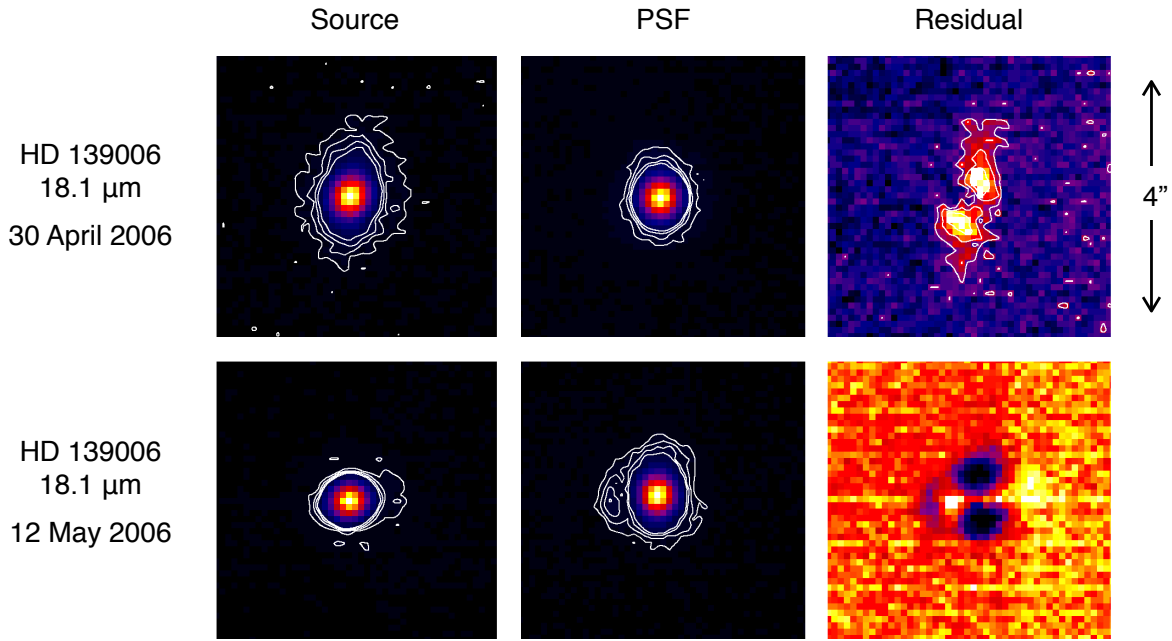}
\caption[HD~139006 images.]{11.2-$\mu$m images of HD~139006, its PSF reference star, and the residual emission following PSF subtraction, where the PSF was scaled to match the peak emission of the disk source.  Contours are drawn at 3-$\sigma_{bkd}$ intervals from 6$\sigma_{bkd}$ to 15$\sigma_{bkd}$.  Note that HD~139006 is visibly elongated compared to the PSF in the first night of data, and less so in the second night of data (in which the PSF also appears elongated).  We believe that these images may suffer from so-called ``chop tails,'' an issue that has since been resolved at Gemini.
\label{fig:139006}}
\end{figure}

\textbf{HD 139006}--
This source is resolved at 11.2~$\mu$m with a statistical significance of 8.6$\sigma_{ext}$.   A Student's $t$-test yields a $p$-value of 2~x~10$^{-7}$, with a 95\% confidence interval that the source FWHM should be broader than the PSF FWHM by 0.041''-- 0.068''.   However, in two sets of 18.1-$\mu$m images from two different nights, one is resolved and one is not.  We believe that this discrepancy arises from image elongation associated with chopping and nodding procedures, which can be seen clearly in a comparison images of HD~139006 at 18.1~$\mu$m on two different nights in Fig.~\ref{fig:139006}.  Indeed, in one set, the PSF profile is broader than that of the source (confirmed by a Student's $t$-test $p$-value of 6~x~10$^{-7}$), and thus we do not claim that the source is spatially resolved at 18.1~$\mu$m.
%We detect no significant excess IR emission in our images of HD~139006.
%The same consideration applies for HD~139006 as was discussed above for HD~75416 and HD~95418.  
While the large photometric uncertainty $\sigma_{phot}$ renders the observed excess emission statistically insignificant, the measurements are well within 2$\sigma_{phot}$ of a formal excess detection, so the results are consistent.

\textbf{HD 181869}--
HD~181869 is resolved at 10.4~$\mu$m (3.1$\sigma_{ext}$) and unresolved at 18.3~$\mu$m.   A Student's $t$-test on the 10.4-$\mu$m data yields a $p$-value of 0.041, and a 95\% confidence interval that the source FWHM should be 0.001''--0.048'' greater than the PSF FWHM.  We have rejected the profile measurement  data at 18.3~$\mu$m on the basis of the PSF being broader than the source, which was confirmed by a Student's $t$-test with a $p$-value of 0.029. No excess emission associated with HD~181869 was detected at either 10.4~$\mu$m or 18.3~$\mu$m, and the simultaneous resolution and lack of detected excess seem contradictory.  However, the measured excess emission at 10.4~$\mu$m is within 2$\sigma_{phot}$ of a marginal excess detection, which is consistent with the apparent spatial resolution at that wavelength.

\section{Discussion}
\label{sec:unresolveddiscussion}

\begin{figure}
\includegraphics[width=\columnwidth]{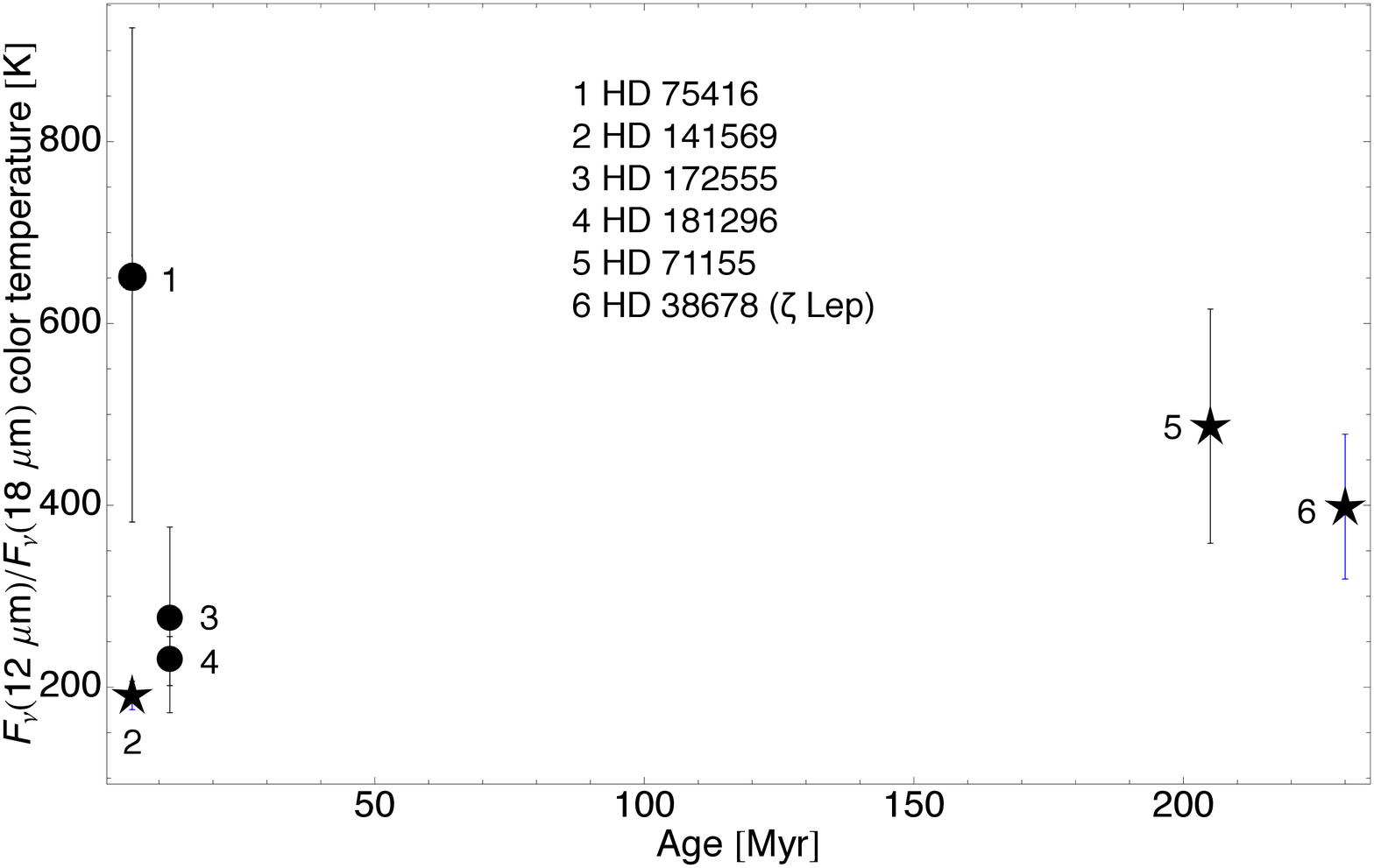}
\caption[Mid-IR color temperature of dust against system age.]{Mid-IR color temperature of dust versus system age.  Age values are the average of all estimates quoted by \citet{rie05}, with the exception of HD~38678 (230~Myr).  Sources represented by filled circles have standard color temperatures as estimated for unresolved sources (see text).  Sources represented by star-shaped points have been spatially resolved by mid-IR images from this study (and in the case of HD~141569, also by prior works).  For reference, $\beta$~Pic has an age of 12~Myr and a color temperature of $\sim$180~K. 
\label{fig:agetempplot}}
\end{figure}

To better understand the nature of the unresolved sources (among those that are still considered debris disk candidates), their colors were used to compare these sources to spatially resolved debris disks whose structure is relatively well known.  In Figure~\ref{fig:agetempplot}, the mid-IR color temperature is plotted against the age for each debris disk candidate.  The resolved sources are represented by star symbols, and the unresolved sources are represented by filled circles.  Most of the sources that have been resolved with ground-based mid-IR imaging observations also have the cooler color temperatures.  We expect cooler dust to be more distant from the star and therefore be part of a more extended disk that is easier to resolve.  Thus, the fact that most of the resolved sources have relatively cool dust is not surprising. 

Of course, our sample is limited in number (eight, after non-disk sources and null excesses have been culled) and biased.  The sources were chosen on the basis of their 24-$\mu$m excess, so it is already known that they have some warm dust, although possibly not hot enough to emit significantly at 12 or 18~$\mu$m, as the null excess detections suggest.  
In addition, it is well known that there are more sources with high fractional luminosities at younger ages, especially less than 20~Myr \citep{rie05,su06,cur08}, and our sample, which was chosen with a brightness criterion, reflects that trend.  
The cluster of sources toward the left of the plot, at ages less than 100~Myr, must be considered with these biases in mind.

Although our sample is not statistically significant, the lack of sources in Figure~\ref{fig:agetempplot} with ages in the range $\sim$50--200~Myr is thought-provoking.  In the cluster of sources at young ages, we have resolved one source with a cool dust temperature that is comparable in extent to the Kuiper Belt (along with $\beta$~Pic, HR~4796A, and HD~32297, as further examples).  In contrast, with the exception of HD~75416 (5~Myr), which has a very large uncertainty in the dust temperature, the two sources in our sample with significantly hotter dust populations are also significantly older: HD~38678 ($\zeta$~Lep) at $\sim$230~Myr and HD~71155 at $\sim$200~Myr.  HD~71155 is a new spatially resolved source with a dust disk radius implied by its 10.4-$\mu$m extent of 2.0~$\pm$~0.8~AU, and this size is also comparable to the size of the solar system's asteroid belt.  Our sample also includes two sources (HD~95418 and HD~139006) that do not have a statistically significant detection of excess IR emission from our data set, but whose spatial extents imply disk radii similar to that of the asteroid belt.  As mentioned in \S\ref{sec:phot_estimates}, the measurements of the excess emission and the spatial extension are not necessarily inconsistent due to the error bars associated with each value.

While low-resolution surveys show that the mid- and far-IR emission from disks generally diminishes with time as the inverse of the system lifetime \citep{rie05, su06}, our observations of two apparent asteroid-belt analogs in our sample imply that somewhat older ($>$ few 100~Myr) sources can sustain significant mid-IR emission above the average levels by ongoing production of dust in asteroid-belt-type collections of planetesimals relatively close to the star.  Whether the collisions have been occurring in a steady state is not obvious, but the amount of the IR excess may help to answer that question.   \citet{wya07b} distinguish disks as having potentially transient dust-producing events if they have greater than 1000 times the maximum fractional IR luminosity predicted for their age that have experienced only steady-state collisions.   In the case of $\zeta$~Lep, the 24-$\mu$m excess exceeds the expected level due to steady-state collisions alone by more than a factor of 10.  However, the excess level for HD~71155 falls within the envelope of expected values for disks experiencing solely steady-state collisions \citep{wya07b}.  Therefore, it may be more plausible (but not imperative) to invoke a process such as delayed stirring or an event analogous to the Earth and moon-progenitor collision for $\zeta$~Lep, whereas observations of HD~71155 seem to be consistent with steady-state evolution.

It is worth noting that the fractional IR luminosities of sources in our sample (with formal detections of excess emission) that we consider to be Kuiper Belt analogs (e.g., HD~32297, HR~4796A; $L_{IR}/L_{\star}\sim$10$^{-3}$) are $\sim$100 times higher than that of sources that we consider to be asteroid belt analogs (e.g., HD~38678 [$\zeta$~Lep], HD~71155; $L_{IR}/L_{\star}\sim$10$^{-5}$).  Thus, in a system that hosts both asteroid-belt-like and Kuiper-Belt-like structures, the presence of a Kuiper Belt with a significantly larger amount of dust may make it difficult to discern the emission from an asteroid belt (see also Liou \& Zook 1999\nocite{lio99}).  However, with the advent of the next generation of ground-based telescopes ($>$30-m), the improvement in diffraction-limited resolving power should enable MIR cameras to distinguish both belts in such systems.

For the remaining unresolved sources that sustain statistically significant IR excesses, what can the presence of the warm dust tell us?  Ultimately, we would like to know the distribution of the dust both radially and azimuthally in order to investigate the planetary system's architecture.  That will hold clues to the production of the dust (e.g. steady-state or catastrophic collisions) and what maintains it in its current location (e.g., shepherding planets).  There are (as of the time of writing) approximately 10 disks that are known to harbor planets (e.g., Wyatt 2008\nocite{wya08}).  It is currently easier to make radial velocity planet detections (the primary detection technique) around FGK-type stars, while it is easier to spatially resolve the thermal emission from  dust disks around the much more luminous A-type stars, and so there is unfortunately little overlap in the detections.  If information about the planetary orbits is known, however, dynamical simulations may indicate where the dust is likely to be stable and how and where the dust was initially produced.  Such simulations have been made for the K-star debris disk HD~69830, which sustains a surprisingly high amount of dust for its 4--10~Gyr age in addition to three Neptune-mass planets \citep{bei05,lov06, lis07}.  \citet{lov06} showed that, given the locations of the planets as determined by radial velocity measurements, there are two stable radii for dust annuli.   The recent direct detection of an orbiting body apparently sculpting the sharp inner edge of the debris disk of Fomalhaut \citep{kal08} highlights this relationship.

In future work, observations at 8-meter facilities with longer integration times and tighter constraints on image quality may reveal more details of the disk structure for some of the sources in this sample, particularly for the ``borderline'' cases.  For example, four sources in our sample (HD~161868, HD~172555, HD~178253, and HD~181296) are not spatially resolved, but the color temperature of their excess IR emission corresponds to that of dust particles emitting like blackbodies in the approximate region of the asteroid belt ($\sim$1--3~AU).
High-resolution imaging at other wavelengths such as the near-IR or submillimeter may also provide a more complete picture of the disk morphology (e.g., Maness et al. 2008\nocite{man08}, Debes et al. 2009\nocite{deb09}, Fitzgerald et al. 2007\nocite{fit07a}).  For disks with especially small angular sizes ($\lesssim$0.1''), interferometric observations in the near-IR and mid-IR have also yielded useful constraints on disk morphologies (e.g., Smith et al. 2009\nocite{smi09b}, Akeson et al. 2009\nocite{ake09}).

\acknowledgments

MMM gratefully acknowledges fellowship support from the Michelson Science Center. This work was performed [in part] under contract with 
JPL funded by NASA through the Michelson Fellowship Program. JPL is managed for NASA by the California Institute of Technology.  This research was partially supported by NSF Grant AST 0098392 to CMT.  Observations were obtained at the Gemini Observatory, operated by AURA, Inc., under agreement with the NSF on behalf of the Gemini partnership: NSF (US), PPARC (UK), NRC (Canada), CONICYT (Chile), ARC (Australia), CNPq (Brazil), and CONICET (Argentina).

{\it Facilities:} \facility{Gemini:North (Michelle)}, \facility{Gemini:South (T-ReCS)}.

\appendix

\section{Appendix: Detailed Profile Width Measurements}

\begin{figure}
\begin{centering}
\includegraphics[width=\columnwidth]{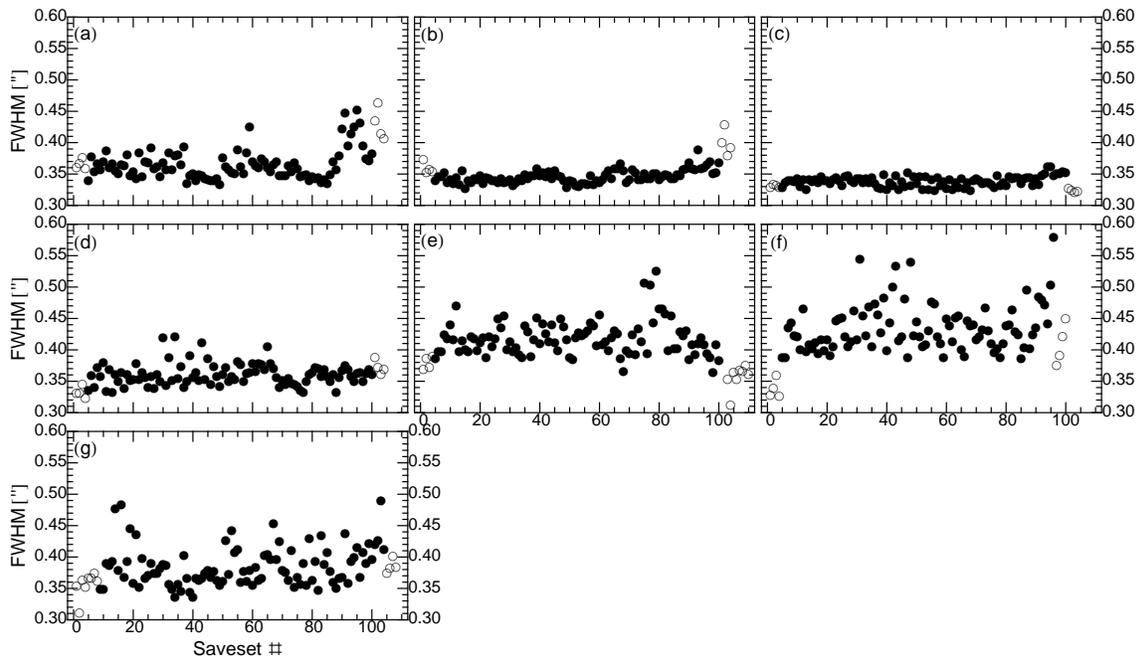}
\caption[N prime sets]{FWHM of profile fits to the sources at 11.2~$\mu$m, per saveset.  Open circles represent the PSF reference star, and filled circles represent the debris disk target. Sources: (a)~HD~56537, (b)~HD~83808, (c)~HD~95418, (d)~HD~102647, (e)~HD~139006, (f)~HD~141569, (g)~HD~161868.
\label{fig:Npsets}}
\end{centering}
\end{figure}

\begin{figure}
\begin{centering}
\includegraphics[width=\columnwidth]{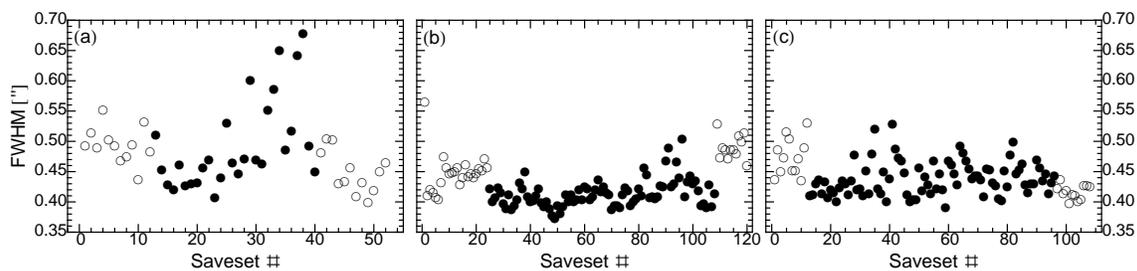}
\caption[Si-5 sets]{FWHM of profile fits to the sources at 11.7~$\mu$m, per saveset.  Open circles represent the PSF reference star, and filled circles represent the debris disk target. Sources: (a)~HD~75416, (b)~HD~115892, (c)~HD~178253.
\label{fig:Si5sets}}
\end{centering}
\end{figure}

\begin{figure}
\begin{centering}
\includegraphics[width=\columnwidth]{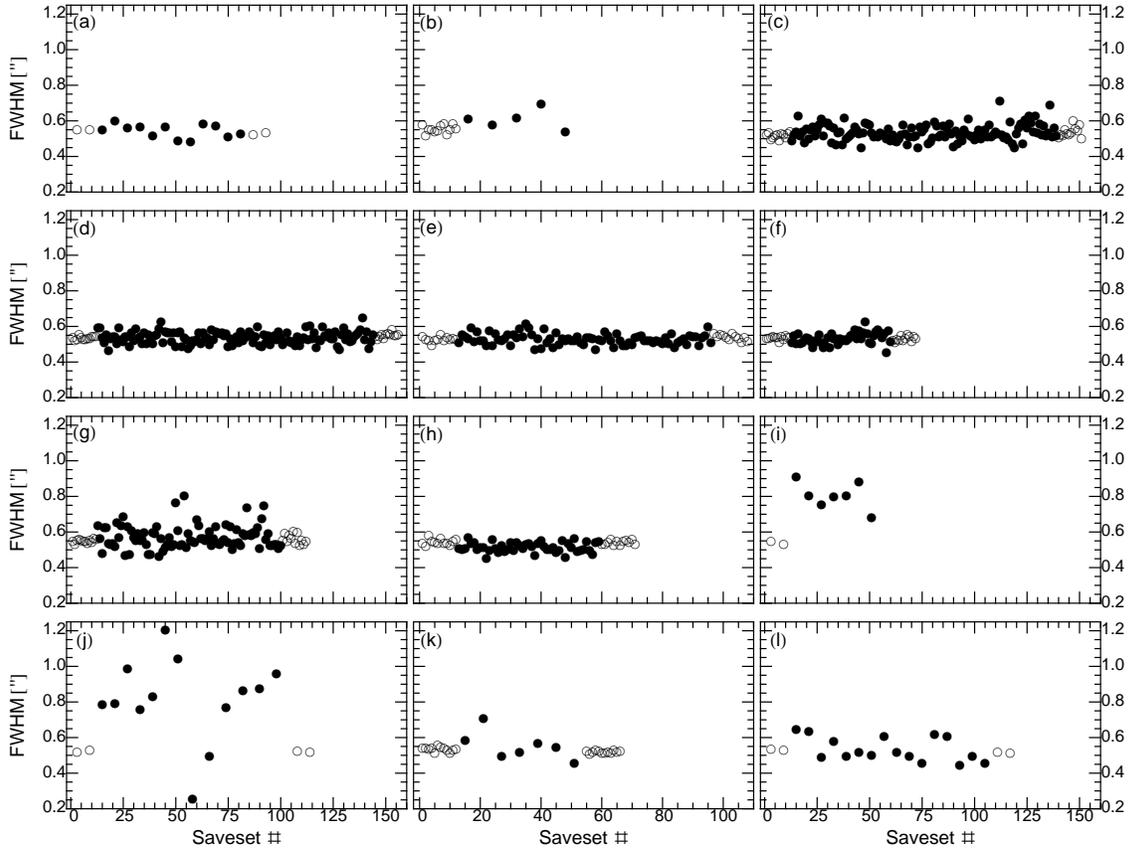}
\caption[Q (Michelle) sets]{FWHM of profile fits to the sources at 18.1~$\mu$m, per saveset.  Open circles represent the PSF reference star, and filled circles represent the debris disk target. Sources: (a)~HD~56537, (b)~HD~56537, (c)~HD~83808, (d)~HD~95418, (e)~HD~102647, (f)~HD~102647, (g)~HD~139006, (h)~HD~139006, (i)~HD~141569, (j)~HD~141569, (k)~HD~161868, (l)~HD~161868.
\label{fig:Qmsets}}
\end{centering}
\end{figure}

\begin{figure}
\begin{centering}
\includegraphics[width=\columnwidth]{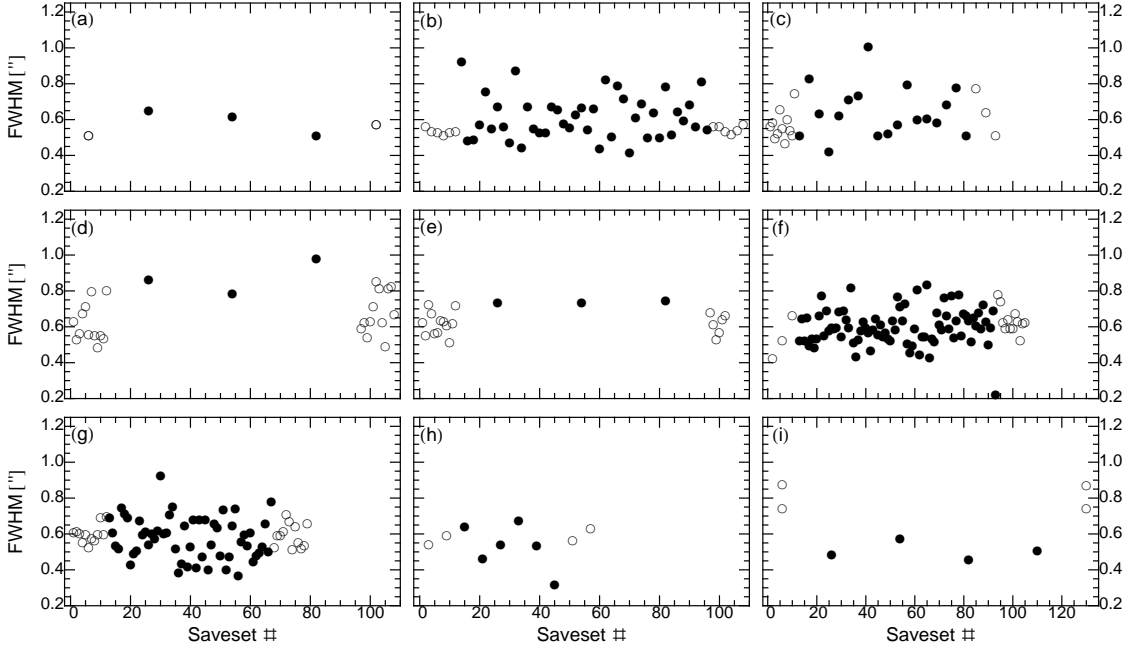}
\caption[Q(T-ReCS) sets]{FWHM of profile fits to the sources at 18.3~$\mu$m, per saveset.  Open circles represent the PSF reference star, and filled circles represent the debris disk target. Sources: (a)~HD~38206, (b)~HD~38678, (c)~HD~71155, (d)~HD~75416, (e)~HD~80950, (f)~HD~115892, (g)~HD~115892, (h)~HD~178253, (i)~HD~181869.
\label{fig:Qtsets}}
\end{centering}
\end{figure}

Here we provide the details of the FWHM measurements of the debris disk candidates and their corresponding PSF reference stars (Figures \ref{fig:Nsets}, \ref{fig:Npsets}, \ref{fig:Si5sets}, \ref{fig:Qmsets}, and \ref{fig:Qtsets}).  As discussed in \S\ref{sec:extent}, the total integration time for an image was broken up into sub-images each corresponding to a fraction of the total time, such that the FWHM of the source could be sampled as frequently as possible.  

When S/N levels allowed, the smallest unit of time for a sub-image was that corresponding to a saveset, $\sim$10~s.  A saveset is a stack of chopped images (on- and off-source), and there are typically three savesets per nod position.  For formal and final image stacking, images from both nod positions must be combined to remove the radiative offset.  However, the S/N of the sources was high enough that the radiative offset did not affect the profile fits to the sources.  Measuring the FWHM in single savesets had the additional benefit of not incorporating positional errors arising from telescope motion.  When images are taken at two nod positions, we expect that the source location is the same in both images.  However, there may be a slight positional inaccuracy occurring between each nod switch, and this is avoided by not combining images from two nod positions.

When the S/N levels were not high enough to perform a reasonable profile fit to the source in a single saveset, these frames were binned up until a sufficient S/N level was reached.  In the following plots, the total number of savesets is shown as a temporal series along the x-axis.  If the FWHM was measured in each saveset image, then the number of data points equals the number of savesets.  If, for example, six savesets had to be binned for a FWHM measurement, then there will only be one data point for every six savesets, and the data point will be shown at the center of the binned saveset group, e.g., savesets 1--6 are binned, so the FWHM value is plotted above the ``saveset \#3'' tick mark.

\begin{figure}
\begin{centering}
\includegraphics[width=\columnwidth]{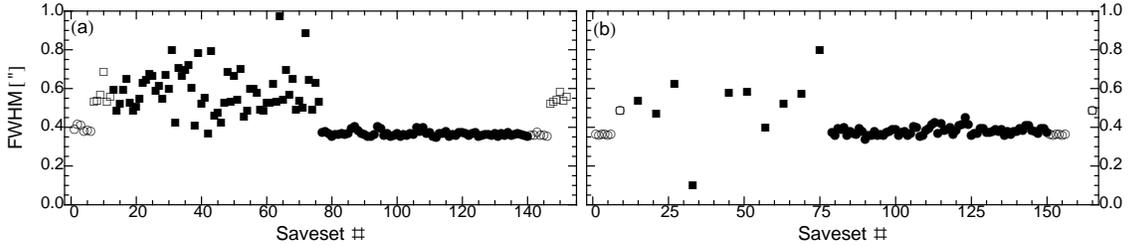}
\caption[Multiple filter sets]{FWHM of profile fits to the sources, per saveset.  Open circles represent the PSF reference star at 11.7~$\mu$m, and filled circles represent the debris disk target at 11.7~$\mu$m.  Open squares represent the PSF reference star at 18.3~$\mu$m, and filled squares represent the debris disk target at 18.3~$\mu$m. Sources: (a)~HD~172555, (b)~HD~181296.
\label{fig:bothsets}}
\end{centering}
\end{figure}

\bibliographystyle{apj}
\bibliography{../moerchen_bibtex}

\clearpage
\end{document}